\documentclass[]{interact}

\usepackage{epstopdf}
\usepackage[caption=false]{subfig}
\usepackage{amsmath}
\usepackage{xcolor}
\usepackage{ulem}
\usepackage[numbers,sort&compress]{natbib}
\bibpunct[, ]{[}{]}{,}{n}{,}{,}
\renewcommand\bibfont{\fontsize{10}{12}\selectfont}
\makeatletter
\def\NAT@def@citea{\def\@citea{\NAT@separator}}
\makeatother

\theoremstyle{plain}

\theoremstyle{definition}

\theoremstyle{remark}

\begin{document}


\title{On spatial beam self-cleaning from the perspective of optical wave thermalization in multimode graded-index fibers}

\author{
\name{Mario Ferraro\textsuperscript{a,b,*}, Fabio Mangini\textsuperscript{a,*}, Mario Zitelli\textsuperscript{a},  and Stefan Wabnitz\textsuperscript{a,c}\thanks{Corresponding author Stefan Wabnitz. Email: stefan.wabnitz@uniroma1.it\\ ${}^*$These authors contributed equally.}}
\affil{\textsuperscript{a}DIET
, Sapienza University of Rome, Via Eudossiana 18, 00184 Rome, Italy\\ \textsuperscript{b}Department of Physics, University of Calabria, Via P. Bucci, 87036 Rende, Italy\\
\textsuperscript{c}CNR-INO, Istituto Nazionale di Ottica, Via Campi Flegrei 34, 80078 Pozzuoli, Italy
}
}

\maketitle

\begin{abstract}
The input power-induced transformation of the transverse intensity profile at the output of graded-index multimode optical fibers from speckles into a bell-shaped beam sitting on a low intensity background is known as spatial beam self-cleaning. Its remarkable properties are the output beam brightness improvement and robustness to fiber bending and squeezing. These properties permit to overcome the limitations of multimode fibers in terms of low output beam quality, which is very promising for a host of technological applications. In this review, we outline recent progress in the understanding of spatial beam self-cleaning, which can be seen as a state of thermal equilibrium in the complex process of modal four-wave mixing. In other words, the associated nonlinear redistribution of the mode powers which ultimately favors the fundamental mode of the fiber can be described in the framework of statistical mechanics applied to the gas of photons populating the fiber modes. On the one hand, this description has been corroborated by a series of experiments by different groups. On the other hand, some open issues still remain, and we offer a perspective for future studies in this emerging and controversial field of research. 
\end{abstract}


\section{Introduction: spatial beam self-cleaning in multimode fibers}

As well known, transporting information over multimode fibers (MMF) suffers from two main limitations. In the time domain, their capacity to transport high-bit-rate signals is hampered by the large dispersion of mode group velocities. In the spatial domain, the capability of MMF to carry images is corrupted by multimode interference, owing to different modal phase velocities. Moreover, the resulting finely speckled output intensity pattern is highly sensitive to any external perturbations such as stress or bending, which lead to mode power transfers via linear mode coupling. 
Because of these reasons, singlemode fibers (SMF) have prevailed for most optical data communication and beam delivery applications. In these fibers only one mode is guided, so that modal dispersion is suppressed and the input laser Gaussian-like beam shape is naturally preserved upon propagation \cite{agrawal2000nonlinear}.
The presence of the third-order nonlinearity or Kerr effect introduces in MMF an additional mechanism of mode coupling via four-wave mixing (FWM), as well as to differential nonlinear phase shifts. 
The resulting interplay of linear and nonlinear mode coupling leads to highly complex, although controllable via the input power and laser beam coupling conditions, spatiotemporal wave dynamics \cite{wright2015controllable, picozzi2015nonlinear,krupa2019multimode}.  

\begin{figure}[!ht]
\centering\includegraphics[width=14.5cm]{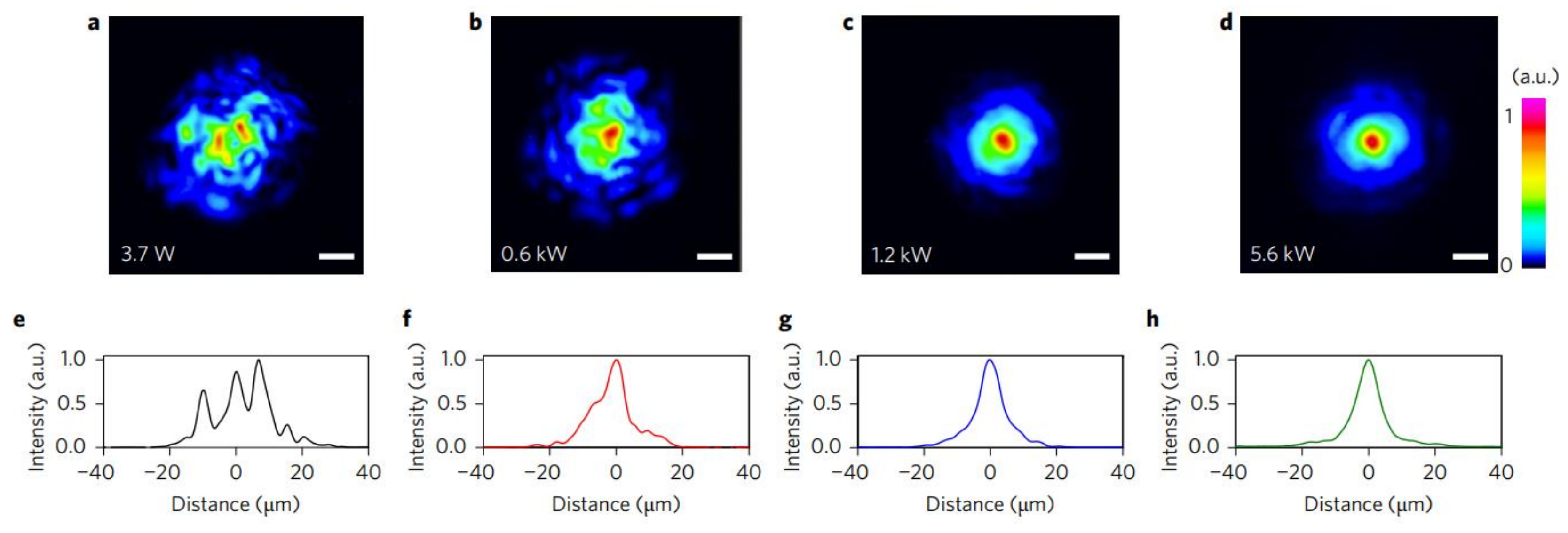}
\caption{Experimental observation of BSC in a 12 m long GRIN MMF. a–d) Near-field images of the MMF output at different input peak powers. The white bars in a-d) are 10 $\mu$m long. e–h) Beam profiles versus x (y = 0 section), corresponding to a-d). [Reproduced with permission from \cite{krupa2017spatial}, [Krupa, Katarzyna, et al., Nature Photonics 11.4 (2017): 237-241.]].}
\label{fig:BSC_original}
\end{figure}

In this context, it was both unexpected and striking to observe that the very Kerr effect could wash out the speckles, and generate to a stable and robust bell-shaped beam at the output of a few meters long graded-index (GRIN) MMF \cite{krupa2016observation,krupa2017spatial}: see Fig. \ref{fig:BSC_original}a-h). In terms of wave dynamics, this spatial beam self-organization provides an example of emergence of order out of complexity, and it is known as beam self-cleaning (BSC)  \cite{krupa2017spatial,liu2016kerr}. Self-induced beam cleanup occurs whenever the input laser power grows larger than a certain threshold value, which depends on fiber length. Specifically, BSC of laser pulses of tens of kilowatts of peak power is experimentally observed over a few meters long GRIN fiber.

\begin{figure}[!ht]
\centering\includegraphics[width=11cm]{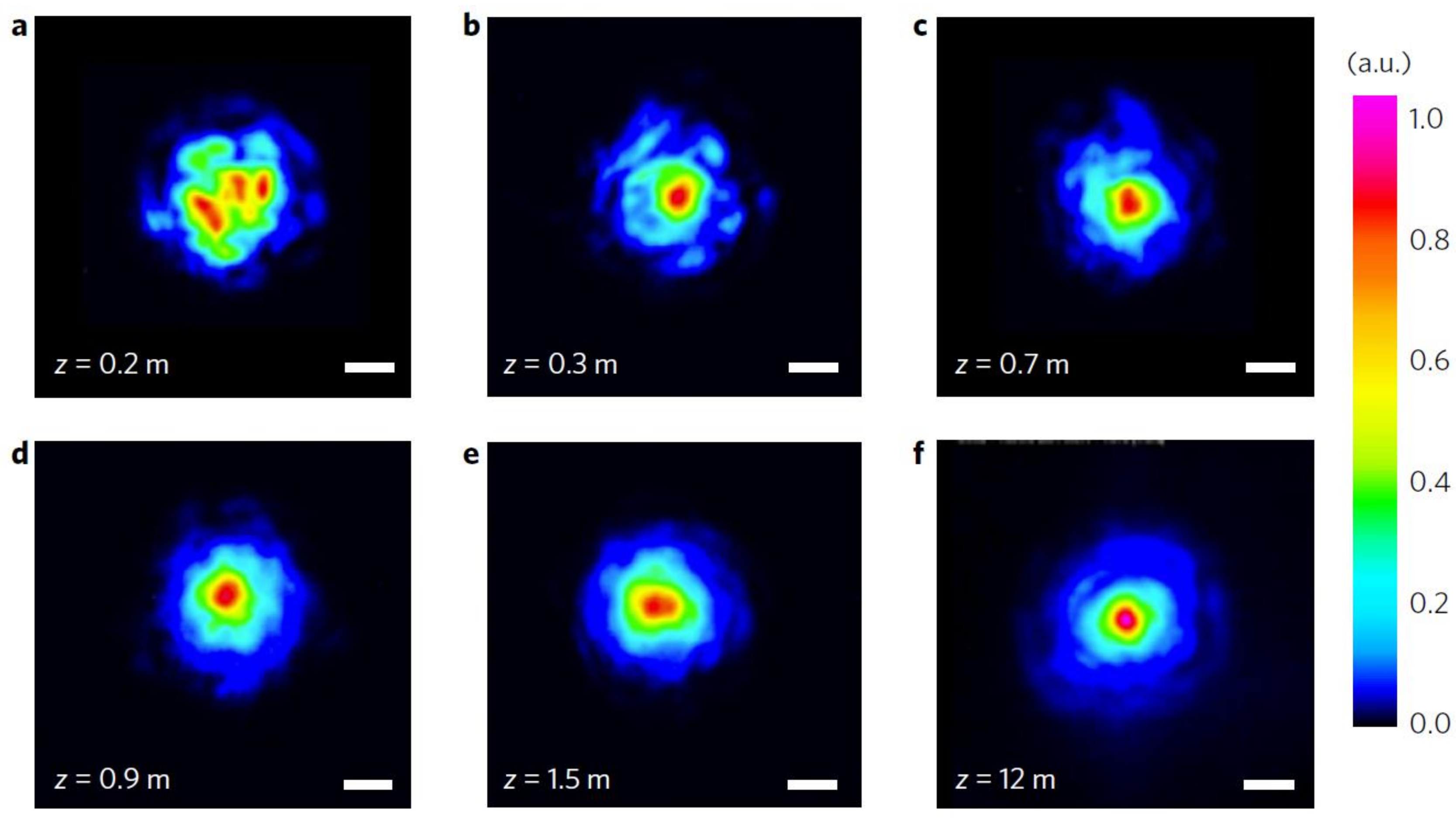}
\caption{\textcolor{black}{Cut-back analysis of BSC for fixed beam power of 44 kW. a–f) Output near-field profile for six different fiber lengths, filtered around the laser source wavelength (1064 nm). [Reproduced with permission from \cite{krupa2017spatial}, [Krupa, Katarzyna, et al., Nature Photonics 11.4 (2017): 237-241.]].}}
\label{fig:BSC_NatPhot2017}
\end{figure}

On the other hand, the results of the fiber cut-back study which are illustrated in Figure \ref{fig:BSC_NatPhot2017}a-f show how beam cleanup can also be obtained by increasing the propagation distance along the GRIN fiber, while keeping the input peak power a constant. This highlights that it is the accumulated nonlinear phase shift, which is proportional to the product of fiber length times power, which determines the occurrence of BSC, rather than a fixed power threshold as it occurs for nonlinear (e.g., Raman or Brillouin) scattering).

The nonlinear reshaping of the output highly multimode beam into a bell-shape is associated with a significant improvement of the beam quality parameter $M^2$, from values, $M^2\simeq 10$ down to  $M^2\simeq 4-1$ \cite{krupa2017spatial,leventoux2020highly,tegin2020single,zitelli2020high}, which is of strong interest for many applications. As a matter of fact, current SMF-based laser beam generation and transport technologies are limited in both pulse peak power and energy. This is due to the SMF small mode size, which sets a relatively low power damage threshold. Whereas, owing to their large mode area, MMF have a greater endurance to high-power laser pulses \cite{ferraro2021femtosecond,ferraro2022multiphoton}. Indeed, the benefit of combining the MMF high power delivery capability without paying a price in terms of beam quality thanks to BSC has already been demonstrated in the context of different laser technologies, from high peak-power mode-locked MMF lasers \cite{tegin2020single,wright2017spatiotemporal} to high-resolution nonlinear microscopy and endoscopy \cite{moussa2021spatiotemporal,wehbi2022continuous}.

In spite of its different experimental demonstrations in GRIN fibers, which have been obtained with laser pulse durations ranging from 1 ns down to 100 femtoseconds, and at wavelengths ranging across the entire transparency window of fibers, from the visible to the infrared region, e.g., in fibers made of soft-glass materials \cite{eslami2022two}
, its physical mechanism is still the subject of a hot debate, and it remains highly controversial in the research community. 

A simple explanation for the power-induced beam cleanup as a result of relatively higher nonlinear losses for the high-order modes (HOM) could be immediately ruled out, since the input/output power transmission remains strictly linear at all intensities involved \cite{krupa2017spatial}. Another possible mechanism of beam cleaning via the loss of spatial coherence resulting from. e.g., nonlinear spectral broadening was also quickly disproved in early experiments. Indeed, at the power threshold for BSC the self-phase-modulation induced broadening remains negligible, in particular for sub-ns pulses \cite{krupa2017spatial}. In addition, self-cleaned beams maintain the spatial coherence of the source, as demonstrated by double slits experiments showing interference fringes resulting from different portions of the same beam \cite{krupa2017spatial}, as well as from two beams, obtained by independently self-cleaning the same laser pulses coming from two different GRIN fibers \cite{fabert2020coherent}.  

Importantly, most experiments reveal that a significant fraction of the output beam energy remains in HOM. When combined  
with the observation that the waist of central bell-shaped beam remains wider than that that of the fundamental mode \cite{leventoux20213d}, one is led to conclude that BSC may be associated with a beam evolution towards a well-defined mode power distribution, rather than simply resulting from a continuous flow of power out of HOM, into the fundamental mode. 


A series of recent experiments has precisely studied the mode power distribution at the output of GRIN fibers, which accompanies BSC. It has been observed that, for self-cleaned beams, the probability of occupation of the fiber modes is closely reproduced by the so-called Rayleigh-Jeans (RJ) law \cite{baudin2020classical,pourbeyram2022direct,mangini2022statistical}. Under certain conditions to be discussed later, this law approximates  the well-known Bose-Einstein distribution for a gas of photons. This important result naturally leads to describing the wave phenomenon of BSC as a termalization of a gas of photons, according to the principles of statistical mechanics. Providing a concise and self-contained overview of the thermodynamic approach to describe nonlinear wave dynamics in MMF is the purpose of this review paper. 

As a matter of fact, a theoretical model based on weak wave turbulence was introduced in 2011 by Ascheri et al., which predicted the occurrence of classical wave condensation as the result of thermalization in a GRIN fiber \cite{aschieri2011condensation}. Subsequently, it was both numerically and experimentally observed that BSC is analogous to 2D hydrodynamic turbulence: the mode power redistribution associated with BSC results from two different mechanisms. Specifically, a flow of energy towards the fundamental mode (known in hydrodynamics as inverse cascade) is accompanied by a simultaneous flow of energy into HOM (direct cascade), at the net expense of intermediate modes \cite{podivilov2019hydrodynamic}.
In 2019, Wu et al. introduced a general description of wave thermalization in highly multimode photon systems, by linking the probabilistic description of their microscopic state with classical macroscopic parameters, such a temperature and chemical potential \cite{wu2019thermodynamic}. In this framework, a multimode beam is described as a gas of particles (photons), which obeys an equation of state at thermal equilibrium. 
Interestingly, both the wave turbulence theory and the gas of particles analogy lead to the same equilibrium distribution of the fiber mode occupation probability, i.e., the RJ law. 

Indeed, both theoretical approaches rely on the same pillars, i.e., the fact that the propagating beam conserves it power ($\mathcal{P}$) and linear momentum, or Hamiltonian ($H$). Typical experimental conditions leading to the observation of BSC fulfill both conservation laws. Linear losses remain negligible, and the powers involved are orders of magnitude lower than the threshold for catastrophic self-focusing, which ensures that the  nonlinear contribution to the Hamiltonian can be neglected with respect to its linear counterpart \cite{agrawal2000nonlinear,ferraro2021femtosecond}. It should be noted, however, that BSC has also been experimentally observed under highly dissipative propagation conditions, e.g., in the presence of heavy loss or strong gain in active MMF \cite{guenard2017kerr,niang2019spatial}. This indicates that BSC is a more robust effect than one could anticipate from thermodynamic descriptions; however extending the analysis of BSC in a dissipative environment goes beyond the scope of this paper.

In this short review, we present a comparative discussion of recent studies, which demonstrate that the mode power distribution associated with BSC may be described as the result of wave thermalization. Our purpose is to make such thermodynamic approach accessible to experimentalists, who are not necessarily closely familiar with the formalism of statistical mechanics. 
To this end, before reporting the experimental results, in Sec. \ref{sec:theory} we retrace the path that leads to the analogy between a multimode laser beam and a gas of particles. This permits us to derive the probability density function for the mode occupation. 
Next, in Sec. \ref{sec:experiments}, we present the main experimental results. Our aim here is to provide the reader with useful recipes on how to link the experiments with theory. For example, how to determine macroscopic thermodynamic parameters (e.g., the beam temperature) from experimental data, and what are the main sources of error in doing so. Finally, we will give our perspective on what are the remaining challenges and open issues. 

\section{The thermodynamic theory in brief}
\label{sec:theory}
Within the thermodynamic theory of nonlinear multimode systems, a self-cleaned beam is as a state of thermal equilibrium for the gas of photons. We may describe this state in terms of just two macroscopic thermodynamic parameters, i.e., its temperature ($T$) and chemical potential ($\mu$). On the other hand, the speckled patterns seen in Fig. \ref{fig:BSC_original}a and b are considered as representing out-of-equilibrium states. 
According to the thermodynamic theory, upon its propagation a multimode beam evolves across successive out-of-equilibrium states, while conserving its optical power ($\mathcal{P}$) and  Hamiltonian ($H$). It is important to underline that BSC is activated by the Kerr nonlinearity or, equivalently, by modal FWM. Therefore, identifying the total beam Hamiltonian with its linear part (or linear momentum) is, a priori, incorrect. This consideration sets the bound of the validity of the thermodynamic approach, which can be only applied in the presence of a weak nonlinearity, i.e., at relatively low powers. This amounts to say that the linear mode structure of the fiber is not perturbed by nonlinearity: this condition would only be violated for powers approaching the threshold for catastrophic self-focusing, that is, at MW power levels. Of course, nonlinearity remains a necessary ingredient for ensuring mode interaction leading to the observed mode power redistribution. Note that, in the absence of nonlinearity, a mode power redistribution may also be introduced il long MMF spans by the presence of linear random mode coupling (RMC), or disorder \cite{olshansky1975mode, zitelli2023spatiotemporal}.

Generally speaking, in physical systems a state of thermal equilibrium is only reached after a critical time, which depends on the system characteristics. In this sense, in the thermodynamic approach to nonlinear beam propagation in MMF, the role of time is played by the propagation length ($z$). As such, experiments such as those illustrated in Figure \ref{fig:BSC_NatPhot2017}a-f
show that a beam progressively self-cleans within a typical fiber distance of a few meters, for tens of kW input power levels. Note that the irreversibility of beam propagation, which is caused by intrinsic perturbations of the fiber leading to RMC, is a key condition for the observation of nonlinearity-induced thermalization. As a matter of fact, the presence of either noise or disorder breaks the time-symmetry, i.e., the reversibility of the propagation equation, which otherwise would conduce to Fermi-Pasta-Ullam-Tsingou (FPUT) recurrences \cite{wabnitz2014instability, vanderhaegen2020observation}.

The thermodynamic theory of BSC only involves the spatial properties of a multimode beam, i.e., it only gives information on the mode occupancy of monochromatic continuous waves. The theory can be derived by statistical mechanics considerations, which lead to a RJ distribution of the mode power fraction at thermal equilibrium. In this section, first we outline the simplest derivation of the RJ distribution for the fiber mode occupation at thermal equilibrium. In order to do so, we recur to the analogy between weakly nonlinear guided waves and the particles of a gas, as discussed in Ref. \cite{makris2020statistical}. 
Next, we are going to derive the equation of state at thermal equilibrium. This is a law that permits to link together different thermodynamic parameters, e.g., the temperature, chemical potential, and the volume. 
Finally, we emphasize the relationship between the predictions of the thermodynamic theories of wave thermalization (and/or condensation) with the experimental demonstrations of BSC.

\subsection{Mathematical derivation of the Rayleigh-Jeans distribution}
\label{sec:subsecMatRJ}

The main idea behind the thermodynamic approach consists of describing a laser beam propagating in an MMF in analogy with a gas of particles (photons), whose average occupation of the multitude of fiber modes irreversibly evolves towards an equilibrium value or distribution.
The number of photons ($N$) is finite, and in the absence of losses it remains fixed upon propagation: it is proportional to the beam power $\mathcal{P}$, i.e., 
\begin{equation}
    N=n_c\mathcal{P},
\end{equation}
where $n_c$ represents the number of photons per unit of power. The number of fiber modes ($M$) is also finite, since it depends on the guiding properties of the MMF: generally speaking $M \propto V^2$, where $V$ is the normalized fiber frequency at the operating wavelength.
We label each mode with an index $i, i=1,2,3...,M$, so that $n_i$ is the number of photons in the $i$-th mode, with propagation constant $\beta_i$. Accordingly, the total number of particles is
\begin{equation}
N=\sum_{i=1}^M n_i,
\label{eq:N-def}
\end{equation}
and the linear part of the Hamiltonian reads as
\begin{equation}
H=\sum_{i=1}^M \beta_i n_i.
\label{eq:H-def}
\end{equation}
Another quantity is conserved upon beam propagation in an MMF, i.e., the longitudinal component of the orbital angular momentum (OAM) of light ($\Omega$), which can be written as
\begin{equation}
\Omega=\sum_{i=1}^M m_i n_i.
\label{eq:L-def}
\end{equation}
Note that, at variance with $N$, $H$, and $\Omega$ which represents constant quantities upon beam propagation, the expressions for $\beta_i$ and $m_i$ depend on the base which is chosen for the linear mode representation. Conserved quantities are what define a \emph{macrostate}, which is associated with macroscopic thermodynamic parameters such as $T$ and $\mu$.

In a statistical mechanics framework, the statistics of a classical gas is described by the Boltzmann distribution, i.e., at thermal equilibrium the probability associated with a given set of mode occupancies $\{n_i\}$, which is referred to as \emph{microstate}, is given by
\begin{equation}
    \rho(\{n_i\}) = \frac{e^{-(aN+bH+c\Omega)}}{\mathcal{Z}},
    \label{eq:rho-def}
\end{equation}
where $\mathcal{Z}$ is the so-called partition function, while $a$, $b$, and $c$ are constants to be determined, associated with the presence of the three invariants $N$, $H$, and $\Omega$, respectively. 
Being a probability function, $\rho(\{n_i\})$ is defined in a way, so that the integral over all microstates which correspond to the same macrostate, or, equivalently, over all possible mode occupations, is equal to 1, i.e.,
\begin{equation}
    \int \rho(\{n_i\}) \prod_{i} d n_i = 1.
    \label{eq:rho-int-1}
\end{equation}
Note that in (\ref{eq:rho-int-1}), which provides an expression for $\mathcal{Z}$, the value of the mode occupancy $n_i$ goes from 0 to infinity. Moreover, it has to be mentioned that the statistical mechanics theory relies on the ergodic hypothesis. This states that macroscopic quantities, e.g., the average occupancy of a fiber mode, can be evaluated as an average over the microstates $\{n_i\}$, instead of an average over ``time". Moreover, it has to be mentioned that the Boltzmann distribution (\ref{eq:rho-def}) is associated with the establishment of thermal equilibrium. Indeed, the Boltzmann distribution provides the maximum value of the entropy ($S$), which is defined as
\begin{equation}
    S = - \int \rho(\{n_i\}) \ln\rho(\{n_i\}) \prod_{i} d n_i,
    \label{eq:S-def-rho}
\end{equation}
and whose expression can be simplified, without any loss of generality, to 
\begin{equation}
   S=\sum_{i=1}^{M}\ln n_i,
   \label{eq:S-ok}
\end{equation}
as it was proposed in Ref. \cite{wu2019thermodynamic}. A demonstration of the equivalence between (\ref{eq:S-def-rho}) and (\ref{eq:S-ok}) can be found in Ref. \cite{makris2020statistical}.

By exploiting the definition of $\rho(\{n_i\})$, the average mode occupancy at thermal equilibrium can be calculated as
\begin{equation}
    \langle n_i \rangle = \int n_i \rho(\{n_j\}) \prod_j d n_j,
\end{equation}
which, after a few algebraic steps, leads to 
the following generalized RJ distribution \cite{podivilov2022thermalization,wu2022thermalization}
\begin{equation}
    n_{i} = \frac{1}{a+b\beta_i+c m_i}.
    \label{eq:RJ-generalized-ab}
\end{equation}
For the sake of readability, we are omitting the average symbol $\langle \cdot \rangle$. 
At this point, it is worth mentioning that most experimental demonstrations of BSC were carried out with beams which do not carry OAM
. In \cite{mangini2022statistical} it was shown that modes with the same value of $\beta_i$ and opposite values of $m_i$ are equally populated. On the other hand, the generalized RJ distribution (\ref{eq:RJ-generalized-ab}) has been recently validated in a recent experiment, where the thermalization of OAM-carrying beams was investigated \cite{podivilov2022thermalization}. For the time being, since our purpose is to compare results obtained by different groups, we restrict our attention to the case where the OAM of light is not taken into account. Thus, by imposing $c$ = 0, we find the RJ law, i.e.,
\begin{equation}
    n_{i} = \frac{1}{a+b\beta_i}.
    \label{eq:RJ-ab}
\end{equation}
Here, we have provided a simple derivation of the RJ law, by imposing that the probability function $\rho(\{n_i\})$ follows the Boltzmann distribution (\ref{eq:rho-def}). 
It should be pointed out that the same generalized RJ equilibrium distribution can be found by recurring to other approaches. For instance, in Ref. \cite{wu2019thermodynamic}, the RJ law for optical multimode systems was retrieved, within a  quantum-like framework, as an approximation of the Bose-Einstein (BE) distribution. Whereas, in Ref. \cite{aschieri2011condensation, podivilov2019hydrodynamic}, the RJ law was obtained by using a weak wave turbulence approach \cite{aschieri2011condensation, podivilov2019hydrodynamic}. The latter is particularly effective for capturing the features of the BSC effect, since it predicts that BSC can only occur in GRIN MMF, whose parabolic refractive index profile greatly facilitates the energy exchange among modes via FWM processes. To the contrary, in step-index MMF, the absence of a coherent superposition of the fiber modes upon beam propagation does not ensure the presence of effectively phase-matched FWM processes. As a matter of fact, BSC was never experimentally observed in step-index fibers, so far (except for the case of few-mode fibers, where the thermodynamic description is meaningless \cite{mohammadzahery2021nonlinear}).


In a GRIN fiber the $M$ modes can be grouped in $Q$ groups of non-degenerate modes. Within each group, to which we associate an index $q$, $q=0,1,2,3,...,Q-1$, the modes can be considered as degenerate, i.e., having the same propagation constant. 
Note that the mode group number $q$ starts from 0, while the modes are usually numbered starting from $i=1$.
According to (\ref{eq:RJ-ab}), degenerate modes with the same propagation constant also have the same probability of occupation at thermal equilibrium (equipartition of the particles belonging to the same group). Therefore, one may write the RJ distribution for the fiber modes as
\begin{equation}
    N_q = \frac{1}{a+b\beta_q},
    \label{eq:RJ-ab-q}
\end{equation}
where $N_q$ is the number of photons per mode in the $q$-th group, to which belong all modes with equal propagation constant $\beta_q$. As such, $\sum_q g_g N_q = N$, where $g_q$ is the group degeneracy.

\begin{figure}[!ht]
\centering\includegraphics[width=10cm]{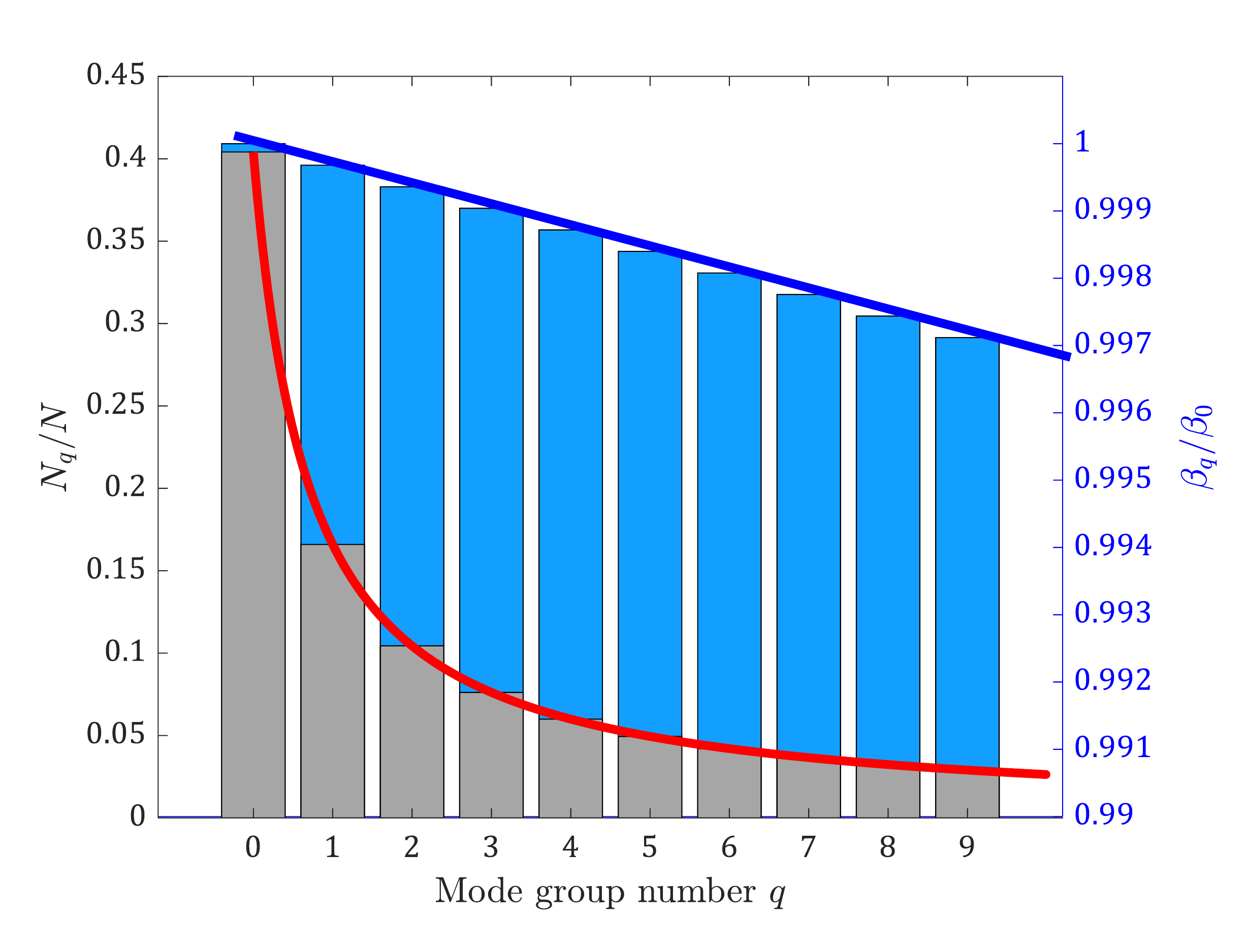}\hspace{5pt}
\caption{Mode occupancy distribution normalized by $N$ at thermal equilibrium in a GRIN MMF (gray histogram) and its mode propagation constants from (\ref{eq:beta-GIF-q}) normalized by $\beta_0$ (light blue histogram) vs. the mode group number $q$. The red and blue solid lines are plot of (\ref{eq:RJ-ab-q}) and (\ref{eq:beta-GIF-q}), treating $q$ as a continuous variable. The parameters used in the plot are $a = \beta_0 + 2$ mm${}^{-1}$, $b$ = -1,  $\Delta$ = 0.0103, and $r_c$ = 50 $\mu$m.
}
\label{fig:GRIN-RJ}
\end{figure}

When decomposed in the Laguerre-Gauss base, the modes of GRIN fibers enjoy the unique property of equally spaced propagation constants, i.e.,
\begin{equation}
    \beta_q = \beta_0 - q \Delta\beta,
    \label{eq:beta-GIF-q}
\end{equation}
where $\Delta\beta = \sqrt{2\Delta}/r_c$ ($\Delta$ and $r_c$ being the core/cladding refractive index difference and the core radius, respectively). Moreover, the mode degeneracy $g_q = 2(q+1)$, where the factor 2 comes from the polarization degeneracy. Here $\beta_0
$ is the largest propagation constant, which characterizes the fundamental mode. Such a ladder structure of the mode propagation constants is shown by means of light blue bars in Fig. \ref{fig:GRIN-RJ}. 

The linear dependence of $\beta_q$ upon $q$ in (\ref{eq:beta-GIF-q}) is illustrated by a blue solid line in Fig. \ref{fig:GRIN-RJ}. Whereas, we show by gray bars the RJ distribution in the mode groups of a GRIN MMF, whose continuous representation in terms (\ref{eq:RJ-ab-q}) is shown by a red solid curve.

 Moreover, since $\beta_0 \gg \Delta\beta$, it is customary to replace $\beta_i$ in (\ref{eq:RJ-ab}) (or $\beta_q$ in (\ref{eq:RJ-ab-q})), with difference between the propagation constant of that mode and that of the highest-order guided mode. Such replacement can be taken into account by redefining the constant parameters $a$ and $b$. This is equivalent to adding an arbitrary additive constant to the linear Hamiltonian, which does not modify the maximization of the entropy, nor the mathematical derivation of the RJ law.
Nevertheless, caution must be exercised when fitting with experimental data. As we will see in the following, in fact, state-of-the-art mode decomposition (MD) methods only allow for determining a limited number of mode groups. Let us suppose that $q=Q'\neq Q-1$ identifies the highest-order mode group that is detectable by the MD. As a result, when comparing different experiments, one has to take into account that different values of $Q'$ lead to different values of the parameters of the RJ distribution, hence, of thermodynamic parameters such as $T$ and $\mu$.

\subsection{The equation of state}

In order to derive the equation of state, which links together the different thermodynamic variables at the point of thermal equilibrium, we need to define the internal energy of the system ($U$): this is proportional to $H$, i.e.,
\begin{equation}
    n_c U = -H.
\end{equation}
In this way, one can write
\begin{equation}
    aN+bH=\sum_{i=1}^M\frac{a}{a+b\beta_i}+\frac{b\beta_i}{a+b\beta_i}=M,
\end{equation}
or, equivalently
\begin{equation}
    an_c\mathcal{P}-bn_cU = M.
\end{equation}
At this point, it is possible to give an interpretation in terms of thermodynamic parameters of the Lagrangian multipliers $a$ and $b$. Specifically, by imposing $an_c=-\mu/T$ and $bn_c=-1/T$, one ends up with an equation of state \cite{wu2019thermodynamic}, i.e.,
\begin{equation}
   U-\mu \mathcal{P}=MT,
   \label{eq:state}
\end{equation}
which allows for determining the values of $T$ and $\mu$ by the mere knowledge of the mode occupancies at the fiber input. As we will see in Sec. \ref{sec:experiments}, this aspect is crucial for the proper experimental validation of the thermodynamic approach.

\subsection{Mode power fractions}
Since in experiments one deals with optical power fractions, it is convenient to define the power carried by the $i$-th mode as $|c_i|^2 = n_i/n_c$, so that the RJ equilibrium distribution reads
\begin{equation}
   |c_i|^2=-\frac{T}{\mu+\beta_i}.
\label{eq:RJ-ok}
\end{equation}
Moreover, it is convenient to write the conservation laws (\ref{eq:N-def}) and (\ref{eq:H-def}) in terms of $|c_i|^2$, i.e.,
\begin{equation}
    \mathcal{P}=\sum_{i=1}^M|c_i|^2,
     \label{eq:P-ci}
\end{equation}
and
\begin{equation}
    U = - \sum_{i=1}^M \beta_i|c_i|^2,
    \label{eq:U-ci}
\end{equation}
respectively. Whereas, the entropy (\ref{eq:S-ok}) can be written as
\begin{equation}
   S=\sum_{i=1}^M\ln |c_i|^2+M\ln n_c.
\label{eq:S-double}
\end{equation}
Since both $M$ and $n_c$ are constant parameters, a more convenient definition of the entropy reads as
\begin{equation}
    S=\sum_{i=1}^M\ln |c_i|^2.
    \label{eq:S-single}
\end{equation}
For more details on the calculation of entropy from experimental data, see Appendix A.

\subsection{Temperature dependence of key parameters}

At this point, it is important to remind that macroscopic thermodynamic parameters, such as $T$ and $\mu$, have a pure statistical meaning: i.e., the photon gas temperature cannot be measured with a thermometer! In our context, the temperature $T$ of a specially prepared multimode beam may even reach negative values \cite{selim2022thermodynamic}: recent experiment have addressed this special propagation regime \cite{baudin2023observation}. Moreover, as it can be easily seen by inspecting Eq. (\ref{eq:RJ-ok}), the RJ distribution has a singularity (i.e., it diverges to infinity), whenever $-\mu$ equals the propagation constant of a given fiber mode. The process of approaching this divergence when the mode involved is the fundamental mode of the GRIN fiber has been referred to as \textit{wave condensation:} in this limit case, the fundamental mode is the only mode with a macroscopic population of photons \cite{aschieri2011condensation,garnier2019wave,baudin2020classical}. Clearly, although BSC may be seen as expressing a tendency of the photon gas to condensate, practical observations of this effect never reach this liming case, except for the trivial case when the input beam is already prepared in a way that the fundamental mode only is excited at the fiber input.

\begin{figure}[!ht]
\centering
\centering\includegraphics[width=11cm]{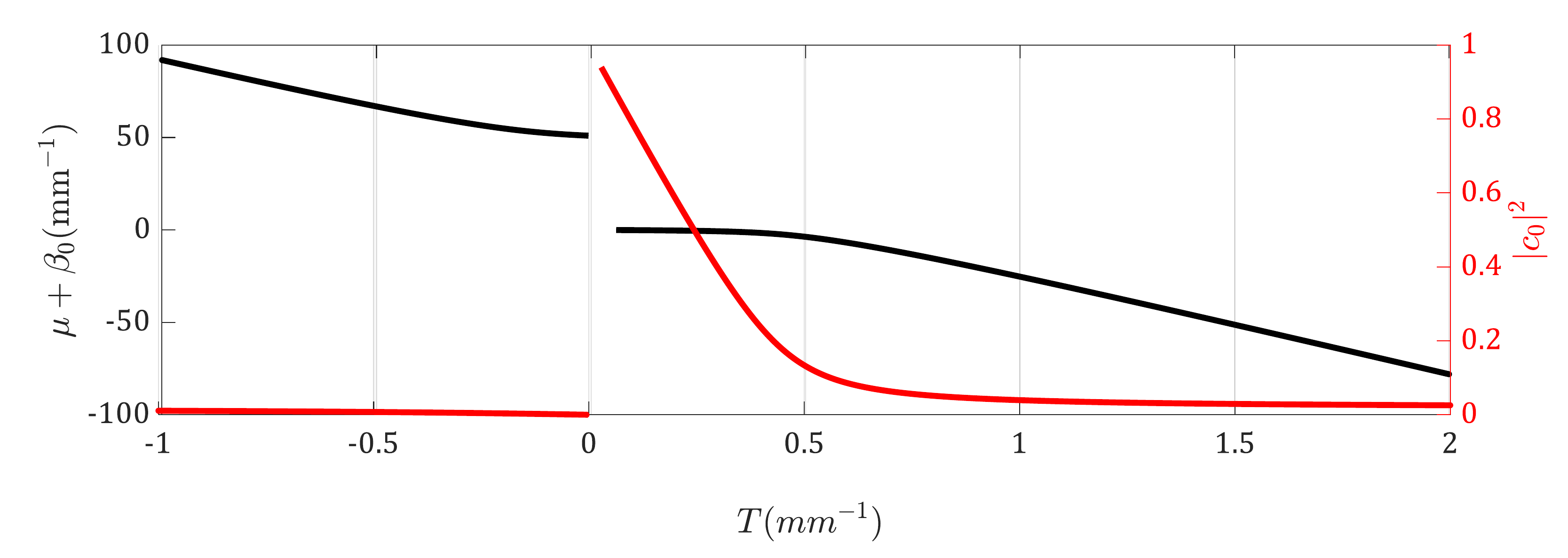}
\caption{Dependence of the thermodynamics parameter on the temperature $T$: (a) chemical potential (black curve) and occupancy of the fundamental mode (red curve); (c) entropy (blue curve) and heat capacity (orange curve). All the parameters are normalized so that $\mathcal{P}$=1.} 
\label{fig:Fabio-RJ}
\end{figure}

In Fig. \ref{fig:Fabio-RJ} we summarize the main properties of a multimode photon gas, in terms of its chemical potential (black curve in Fig. \ref{fig:Fabio-RJ}a) and power fraction in the fundamental mode (red curve in Fig. \ref{fig:Fabio-RJ}a), as a function of its temperature. All plots are obtained by combining equations (\ref{eq:RJ-ok}) and (\ref{eq:state}), and taking into account of the physical condition $|c_i|^2\ge0$. Here, we used the same propagation constant values as in Fig. \ref{fig:GRIN-RJ}, and we normalized all quantities to the beam power, i.e., we set $\mathcal{P}=1$.

Fig. \ref{fig:Fabio-RJ}a shows that, whenever the temperature $|T| \xrightarrow{} 0$, the chemical potential tends to a fixed value, which depends on the sign of $T$. The critical value of zero temperature, in fact, can be reached from either the positive or the negative side the horizontal axis in Fig. \ref{fig:Fabio-RJ}a. In the former case, the occupancy of the fundamental mode tends to unity (condensation). Whereas for negative temperatures the fundamental mode is progressively depleted as $T$ approaches zero. This corresponds to an opposite tendency with respect to the case of BSC, where the fundamental mode is always the most populated in the equilibrium mode power distribution. Note that in Fig. \ref{fig:Fabio-RJ} $\mu$ seems to have a discontinuity in $T$ = 0. Moreover, the curve  $\mu + \beta_0$ vs. T (black curve in Fig. \ref{fig:Fabio-RJ}) is not resolved when $T \sim 0$. This is because, according to the RJ law (\ref{eq:RJ-ok}), the mode power fraction diverges whenever $-\mu$ tends to any $\beta_i$. Therefore when $-\mu \in [\beta_Q,\beta_0]$ (which is the gap of the black curve in Fig. \ref{fig:Fabio-RJ}), $T$ is not associated with an unique value of $\mu$. It must be mentioned, however, that in practical demonstrations of BSC, one has $\mu + \beta_0 \leq 0$ \cite{mangini2022statistical,pourbeyram2022direct,baudin2020classical}.



\subsection{A global thermodynamic perspective of beam self-cleaning}

As discussed in the introduction, the generally accepted definition of BSC is the input power-induced reshaping of the output transverse intensity profile emerging out of a length of multimode fiber, from fluctuating speckles to a robust bell-shape. In the absence of dissipative effects, the only mechanism that can lead to such a reshaping is the energy transfer among modes via FWM interactions. The fact that BSC is facilitated in GRIN fibers can be explained by the presence of beam self-imaging in these fibers, a direct result of their equispaced mode propagation constants \cite{krupa2016observation}. In addition to phase-matching the input laser beam with spectrally distant sidebands growing from noise (an effect that has been called \textit{geometric parametric instability}), the beam intensity oscillations due to self-imaging lead, via the Kerr effect, to a periodic dynamic (or light activated) grating, which may phase-match FWM processes, hence it will greatly enhance the mode redistribution process. 

Now, in a lossless fiber these processes lead to a periodic or recurrent energy exchange for each FWM process taken individually. Therefore, one remains with the question, on how a stable or thermal equilibrium state can ever be reached, where now energy is flowing back and forth with equal probability among the modes, so that no net energy transfer is observed on average.
As a matter of fact, the very early FPUT experiments had surprisingly defied the common sense of their time, which expected a rapid thermalization to emerge in the evolution of chaotic (i.e., non integrable) highly dimensional multimode dynamical systems.
Nevertheless, recent progress in computing capabilities have shown that even the FPUT system eventually thermalizes, albeit after a surprisingly long observation time.

Returning to MMF, there is no question that their evolution is ruled by a chaotic dynamical system, hence that thermalization is eventually to be expected: the only issue being, after how long propagation distance, or, in some sense equivalently, for what strength of the nonlinearity. In this respect, as previously mentioned, the unavoidable presence of RMC or noise will largely determine, in practical experiments, the actual threshold for thermalization to occur. 

These considerations shift the attention from the question of what is the threshold for thermalization, something that any theory would likely be incapable to precisely determine, given that is it intrinsically determined by disorder, to answering the question about how the thermal state looks like, once that it is established. This is indeed the scope of a thermodynamic approach, which inherently deals with transitions between different states of equilibrium of matter, and not with out-of-equilibrium states that approach them.

\begin{figure}[!ht]
\centering
\includegraphics[width=10cm]{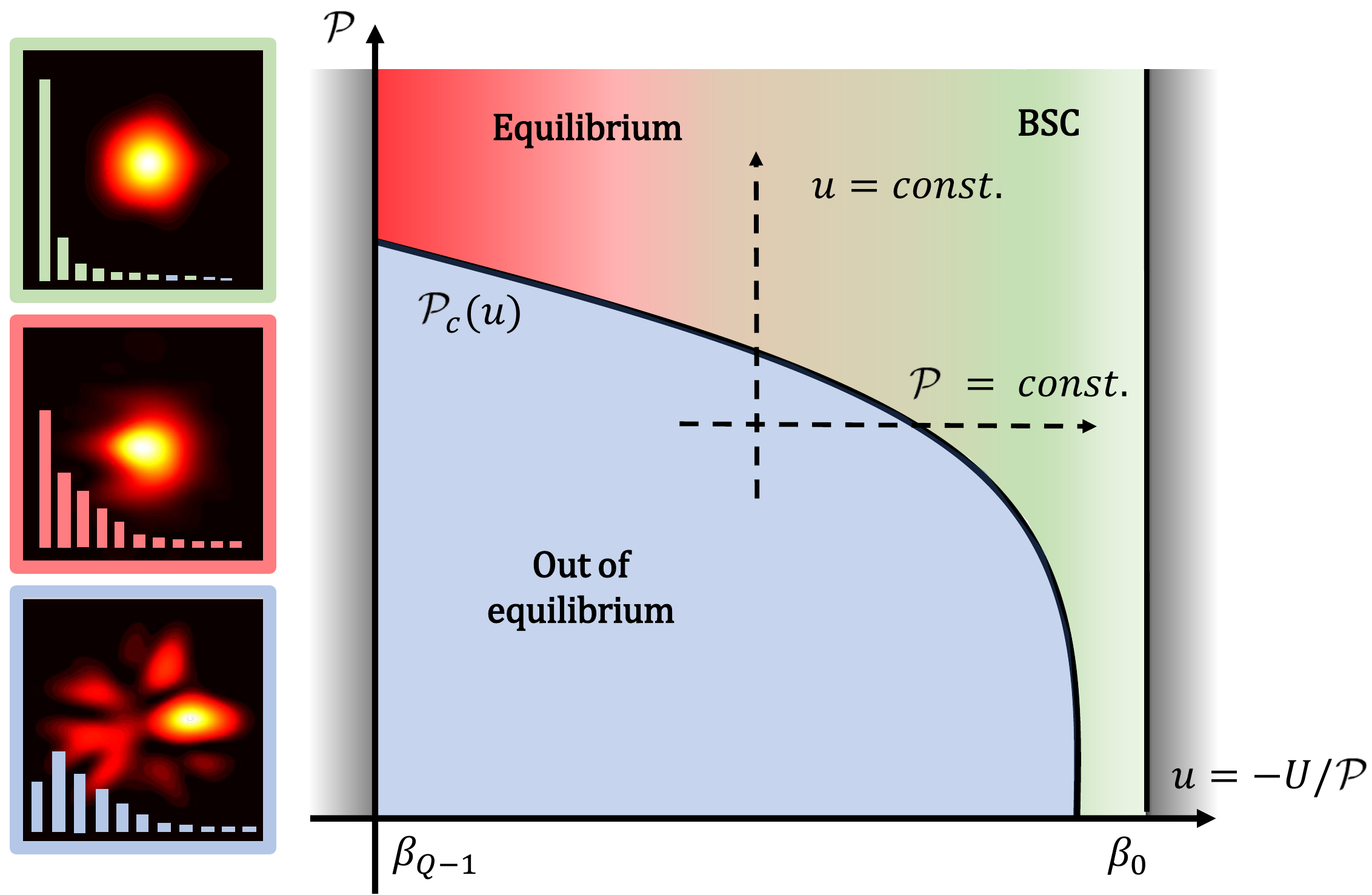}	
\caption{Optical phase diagram describing wave thermalization in a finite length of MMF. The vertical axis indicates the optical power $\mathcal{P}$, and the horizontal axis measures and the internal energy, normalized by power, $U/\mathcal{P}$.
The solid black line (whose exact position is somewhat arbitrary) separates states of thermal equilibrium from out-of-equilibrium states (which populate the light blue region). Among thermalized states, we may define as \textit{clean beam states}, those states where the fractional population of the fundamental mode exceeds a certain threshold (upper-right (green) region). Note that $U/\mathcal{P}$ is fully determined by fixing the input laser beam coupling condition into the MMF. 
The two dashed arrows indicate two orthogonal paths for crossing the 
solid line, i.e., for wave thermalization. The vertical arrow corresponds to increasing the input power: this is done in BSC experiments.
Moving along the horizontal arrow means keeping the input power constant while changing the input coupling so that less modes are progressively excited at the output: this route has been referred to in the literature as thermalization via \textit{condensation}. 
In the upper-right (red) region of the diagram, the beam is at thermal equilibrium, but generally, it does not have a bell-shape, because of the high population of HOM. Images on the inset show examples of output beam intensities in the green, red and blue regions, along with their corresponding mode power distributions.}
\label{fig:diagram}%
\end{figure}

That said, it proves helpful for the understanding of recent experiments dealing with wave thermalization in MMF, to resort to a \textit{phase plane} visualization of the different regions comprising either out-of-equilibrium and equilibrium (or thermalized) states for a multimode fiber system of finite length. In such a diagram, which is schematically shown
in Fig. \ref{fig:diagram}, vertical and horizontal axes correspond to the two quantities that are conserved upon propagation, namely the optical power $\mathcal{P}$ and the normalized internal energy $U/\mathcal{P}$, respectively.

Note that the laser coupling conditions at the fiber input unequivocally determine the value of $u\equiv -U/\mathcal{P}$. In essence, the value of $u$ measures how many modes are excited at the fiber input: larger values of $u$ correspond to less modes, until the fundamental mode only is populated. Hence, the values of $u$ range between a maximum of $\beta_0$, which is attained whenever only the fundamental is excited, to a minimum of $\beta_{Q-1}$, corresponding to the excitation of modes belonging to the largest HOM group. 
The diagram of Fig. \ref{fig:diagram} is split among two regions, which identify equilibrium and out-of-equilibrium beams at the fiber output, respectively. On the bottom left part (light blue area), i.e., at relatively low powers and highly multimode input beams, the output transverse intensity profile remains speckled (light blue framed intensity pattern inset). Indeed, if either the input power is too low, or the input mode occupancy is too far from the equilibrium RJ distribution, no wave thermalization may occur over the given fiber length.

To the contrary, in the green area in Fig. \ref{fig:diagram} is obtained for sufficiently high values of $u$ and/or input powers above the BSC threshold ($\mathcal{P}_c$), one observes a bell-shaped beam at the fiber output (green framed inset in Fig. \ref{fig:diagram}). The separatrix curve between the two regions of the diagram (solid line in Fig. \ref{fig:diagram}) describes the dependence of the critical power on the number of excited modes. Although the exact shape of this separatrix remains elusive, recent experiments reported in Ref.\cite{ferraro2023multimode} have shown that $\mathcal{P}_c$ grows larger exponentially with the incidence angle of the input beam into a GRIN MMF, which is roughly proportional to the number of excited modes. Such a trend was confirmed in experiments of BCS in the anomalous dispersion regime, carried out with a few-mode input beam, showing a dramatic decrease by at least two orders of magnitude of the BSC power threshold \cite{leventoux2020highly}, when compared with earlier experiments involving highly multimode input beams \cite{krupa2017spatial}.

We emphasize that crossing the separatrix in Fig. \ref{fig:diagram} does not provide an abrupt change of the beam shape. To the contrary, the transformation from speckles to a bell-shape takes place gradually. Moreover, 
it has to be kept in mind that the whole phase diagram varies with the fiber length, e.g., the power threshold for triggering the BSC effect decreases as the fiber length grows larger \cite{deliancourt2019kerr}. Accordingly, the area of the blue region in the diagram in Fig. \ref{fig:diagram} progressively quenches when increasing the fiber length. Eventually, at virtually infinite propagation distances, all beams would reach their thermal equilibrium. Thus, the diagram in Fig. \ref{fig:diagram} would only contain the red and the green regions of thermalized beams.

Within the diagram in Fig. \ref{fig:diagram}, the transition from an out-of-equilibrium state (light blue area) into thermalized one, and in particular a stable bell-shaped beam (green area), may occur by varying either the input power or the input coupling conditions. By definition, thermalization leading to BSC is obtained in experiments as a  consequence of an input power increase, with fixed coupling conditions ($u$ = const.), see the vertical arrow in Fig. \ref{fig:diagram}. Conversely, whenever the input power is fixed, one may adjust the input coupling conditions, or number of excited modes, until thermalization is achieved at the end of the fiber. This route to thermalization has been referred to as a \textit{classical wave condensation} process \cite{baudin2020classical}, see the horizontal arrow in in Fig. \ref{fig:diagram}.


However, this terminology appears to be questionable, since the process must not be confused with the condensation phenomenon at $T$=0, that we described in the former section (see Fig. \ref{fig:Fabio-RJ}). 
This requires the cooling of a beam which is already at thermal equilibrium.
On the other hand, in the route to thermalization of \cite{baudin2020classical} the input coupling conditions were varied, leading to a series of out-of-equlibrium states. 
 As a result, the variation of $T$ is a consequence of the variation of $U$, and it is not driven by a source of heat.


We underline that in the theory described in the former Section both $\mathcal{P}$ and $U$ are fixed upon propagation. This is because the constrained maximization of entropy only leads to termalization at some point along the propagation distance (which is the system evolution variable), which in principle might be very large. Whereas, the intensity patterns in Fig. \ref{fig:diagram} are obtained at the output of a fiber of relatively short length.
This means that the route to thermalization leading to BSC in Fig. \ref{fig:diagram} involves passing through different out-of-equilibrium states via an increase the input power in the experiments \cite{krupa2017spatial, mangini2022statistical,pourbeyram2022direct}. 

In other words, thermalization is a physical process which leads to the establishment of an equilibrium starting from an out-of-equilibrium state. According to the theory, such an out-of-equilibrium state lives at the fiber input, whereas in the experiments one usually refers to the beam shape at the fiber output (which is the one shown on the left side of the diagram in Fig. \ref{fig:diagram}). Therefore, the transformation of a speckled output pattern into a bell-shape via the increase of the power may also be referred to as a wave thermalization process. 

In the latter, the role of input power is that of measuring the strength mode interactions via FWM. In this sense, increasing the power can be seen as a way for reducing the propagation distance which is needed to reach thermal equilibrium. 
Having said that, it has to be noted that a bell-shaped beam can only be obtained if the statistical temperature of the beam is sufficiently low. Since the pair ($\mathcal{P}$, $U$) (or ($H$,$N$)) uniquely determines the pair ($T$,$\mu$), a low temperature means favourable enough injection conditions (or a sufficiently low number of excited modes) for a given power.
In this sense, the value of $u$ permits to discriminate whether or not a beam has termalized, for a given input power and fiber length. Generally speaking, at thermal equilibrium, a higher value of $u$ will result in a larger population of the fundamental mode. As a matter of fact, if $u$ is too low, one may still reach thermal equilibrium at the fiber output, but the associated RJ distribution has such a high content of HOM that the resulting beam no longer \textit{looks clean}, i.e., it lacks a bell-shape (see pink region and associated inset in Fig. \ref{fig:diagram}) 
On the other hand, high-quality beams are obtained for sufficiently high values of $u$. There is not a consensus about the definition of the critical value of $u$, that discriminates between bell-shaped and non-bell-shaped beams. This is of course related to the previously discussed  definition of the black separatrix curve in Fig. \ref{fig:diagram}). Yet, in Ref. \cite{baudin2020classical} a critical value of $u$, say $u_c$, was introduced, in order to discriminate whether the beam is sufficiently close to a clean beam.
In the next section, we are going to overview the main numerical and experimental demonstrations of wave thermalization processes in GRIN fibers.

\subsection{Numerical demonstrations}
\label{sec:numerical_model}

\begin{figure}[!ht]
\centering
\centering\includegraphics[width=9cm]{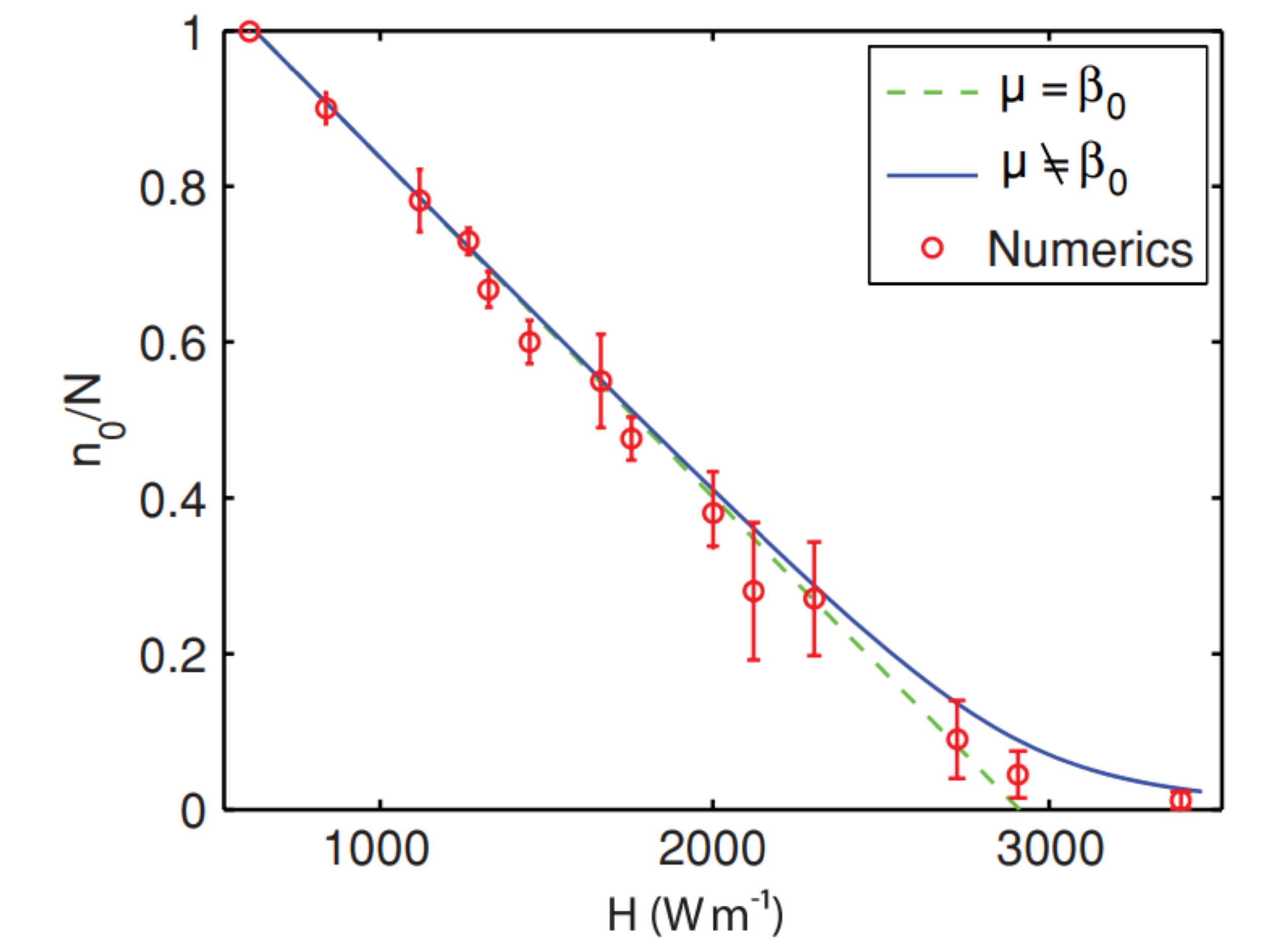}\hspace{5pt}
\caption{Occupancy of the fundamental mode, normalized by the total number of particles $n_0/N$, vs. the Hamiltonian $H$. Red points are results of numerical simulations based on ssolving the 2D NLSE. Error bars for numerical simulations (red dots) denote the amount of fluctuations (standard deviation) of $n_0/N$, once equilibrium is reached. Dashed green and solid blue lines refer to the case of beam condensation and thermalization, respectively. [Reproduced with permission from \cite{aschieri2011condensation}, [Aschieri, P., et al., Physical Review A 83.3 (2011): 033838.]]. Note that the Hamiltonian $H$ is expressed in Wm${}^-1$, since in Ref. \cite{aschieri2011condensation} $H$ is defined as the opposite of the internal energy $U$ in (\ref{eq:U-ci}).} 
\label{fig:sim-condensation}
\end{figure}

\begin{figure}[!ht]
\centering
\centering\includegraphics[width=9cm]{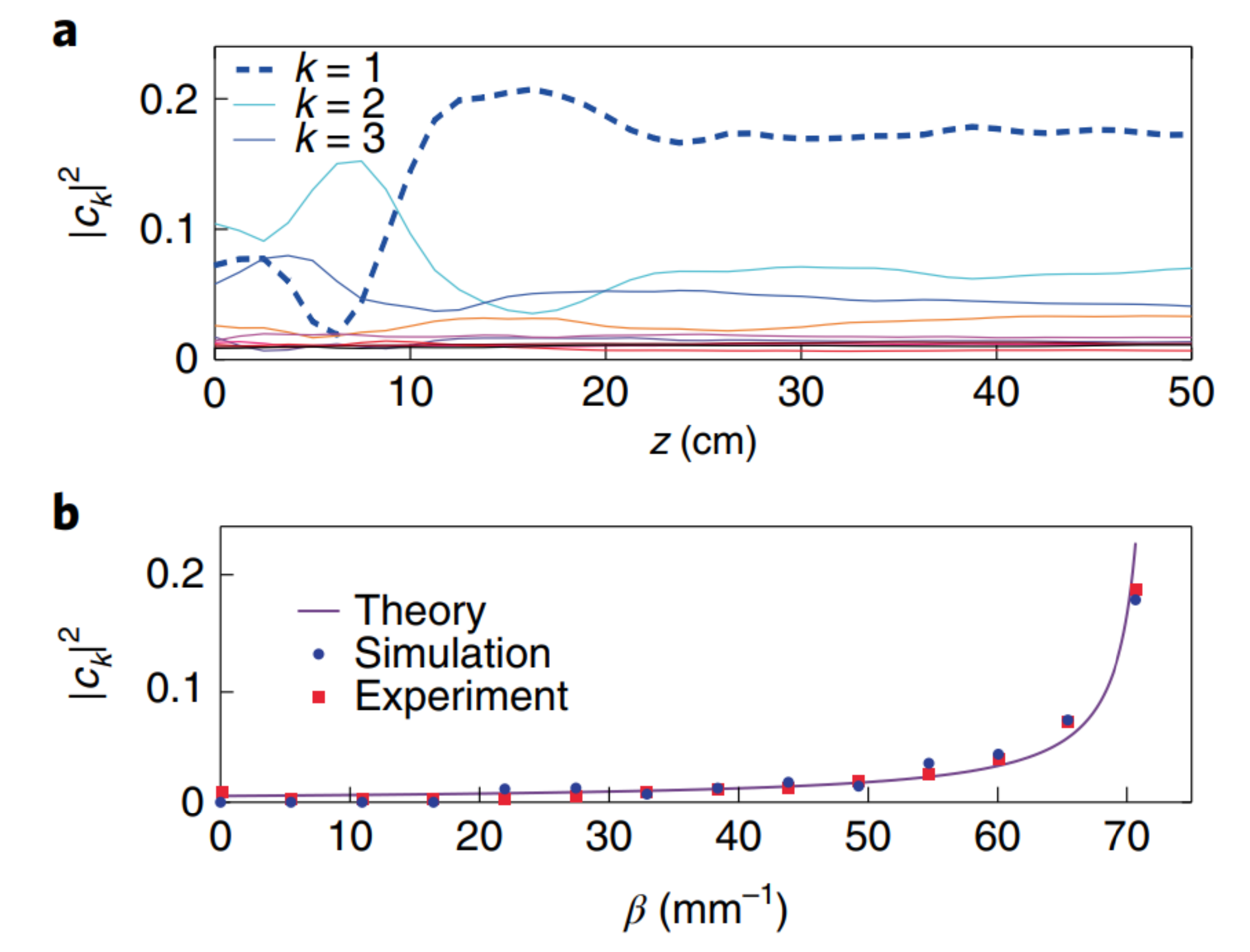}\hspace{5pt}
\caption{Numerical simulations based on coupled-mode equations, describing the optical thermalization process leading to an RJ distribution. a) Mode power fraction in each degenerate group $k$. The simulations consider pulses of 200 fs with 52 kW input power, which propagate over 50 cm of GRIN MMF. Simulations involve the first 55 modes, i.e., $k<10$. b) Comparison of umerical results (blue dots) with the theoretically predicted RJ distribution (solid blue line). The figure also shows the agreement with experimental results, which are discussed in Sec. \ref{sec:experiments}. , [Reproduced with permission from \cite{pourbeyram2022direct}, [Pourbeyram, Hamed, et al., Nature Physics 18.6 (2022): 685-690.]].} 
\label{fig:sim-thermalization}
\end{figure}

The predictions of the thermodynamic approach for describing wave dynamics in GRIN fibers have been confirmed by numerical simulations based on different models, involving the multi-dimensional nonlinear Schr\"odinger equation (NLSE), or coupled-mode equations. In Fig. \ref{fig:sim-condensation} and Fig. \ref{fig:sim-thermalization}, we report results from Ref. \cite{aschieri2011condensation} and Ref. \cite{pourbeyram2022direct}, respectively. For the results of Fig. \ref{fig:sim-condensation}, the underlying model is the purely spatial NLSE which accounts for diffraction, a truncated parabolic refractive index (i.e., that of a GRIN MMF), and Kerr nonlinearity. As can be seen, there is an excellent agreement between numerics (red dots in Fig. \ref{fig:sim-condensation}), and the analytical RJ distribution (solid and dashed lines in Fig. \ref{fig:sim-condensation}), as far as the relative occupation of the fundamental mode is concerned. Here we show the normalized occupancy of the fundamental mode, i.e., $n_0/N$ (=$N_0/N$, since the fundamental mode is non-degenerate), vs. the Hamiltonian $H$. The dashed green curve represents the power condensed in the fundamental mode at $T = 0$, i.e., at $\mu = \beta_0$, in agreement with the black curve in Fig. \ref{fig:GRIN-RJ}a at $\mu + \beta_0 \xrightarrow{} 0$ (note that in Ref. \cite{aschieri2011condensation} the chemical potential is defined with an opposite sign with respect to here). Whereas, the solid blue line corresponds to the case of $T\neq 0$.

Numerical simulations describing thermalization of the mode power distribution according to the RJ law in GRIN MMF were also been reported in \cite{pourbeyram2022direct}. In that work, the Authors solved a system of coupled NLSEs for the temporal envelopes of the spatial modes. The equations include Kerr and Raman nonlinearities, self-steepening, as well as chromatic and modal dispersions. A total of 55 transverse spatial modes of the fibre were taken into account in the model, i.e., all modes with $q<10$ ($\beta_9$ = 20 mm${}^{-1}$). The results of simulations are reported in Fig. \ref{fig:sim-thermalization}. Specifically, Fig. \ref{fig:sim-thermalization}a shows the evolution of the normalized mode power fraction vs. propagation distance ($z$). Note that in Fig. \ref{fig:sim-thermalization}, the symbol $|c_k|^2$ is equivalent to $N_q/N$ with $k=q+1$, and it must not be confused with $|c_i|^2$, since the index $k$ runs over the mode groups, and not over the single modes. As it can be seen, the mode power fraction reaches a steady value after a few tens of centimeters of propagation distance. In particular, the limit values of the mode occupancy were found to be in good agreement with the theoretical predictions of the RJ law, as it is shown in Fig. \ref{fig:sim-thermalization}b.

\section{Experimental demonstrations of wave thermalization}
\label{sec:experiments}

\subsection{Thermalization by varying input power}
\label{sec:exp-varying-N}

Direct experimental demonstrations of BSC as a result of wave thermalization were reported in Ref. \cite{pourbeyram2022direct} and \cite{mangini2022statistical}. In order to measure the output mode power composition from GRIN fibers, both groups used holographic mode decomposition (MD) methods. In Ref.\cite{pourbeyram2022direct}, an interferometric system was used; whereas in Ref.\cite{mangini2022statistical} a phase-only liquid crystal spatial light modulator was exploited (see Ref. \cite{gervaziev2020mode} for details). Besides the different MD technique, the overall results of both studies are rather similar. The good agreement between the theoretical RJ distribution and the experimentally measured mode power fractions, as obtained in Ref.\cite{pourbeyram2022direct}, is shown in Fig. \ref{fig:sim-thermalization}b. This confirms the numerical predictions of Fig. \ref{fig:sim-thermalization}, showing that the RJ distribution is well approached at the fiber output for a given input condition, i.e., with fixed values of $\mathcal{P}$ and $U$.

\begin{figure}[ht!]
\centering\includegraphics[width=14.5cm]{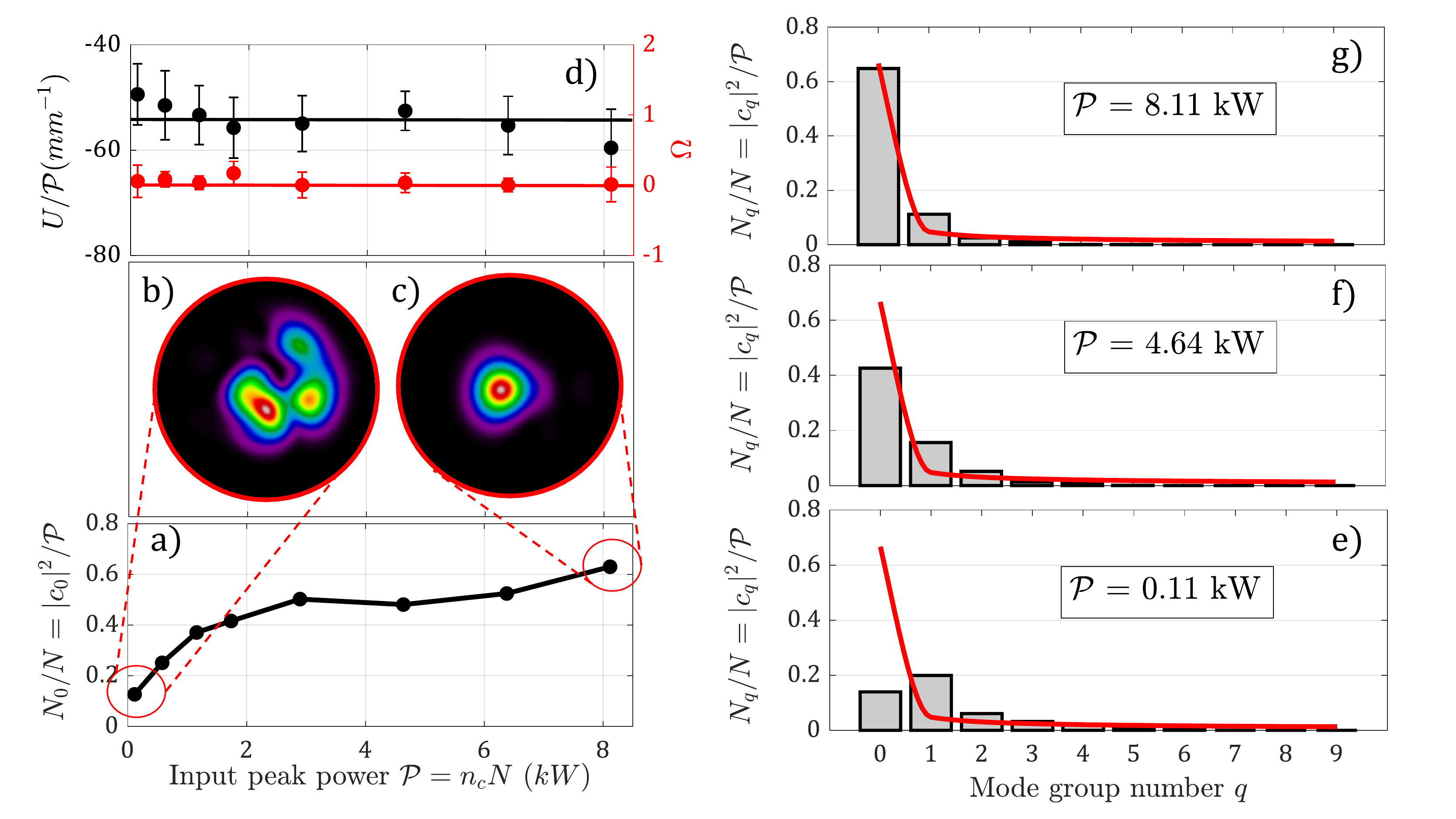}
\caption{Experimental demonstration of BSC as a wave thermalization process. a) Power fraction of the fundamental mode vs. input peak power. b,c) Output beam intensity profile at the lowest and highest values of input peak power, i.e., 0.11 and 8.11 kW, respectively. d) Normalized internal energy $U/\mathcal{P}$ and angular momentum $\Omega$ conservation. The error bars are associated to the accuracy of the mode decomposition method (see Ref. \cite{mangini2022statistical} for details).
e-g) Mode power fraction per group vs. group number (bars) and expected RJ distribution (red line) at 0.11 (d), 4.64 (e), and 8.11 kW (f) of input peak power, respectively. Data are from Ref. \cite{mangini2022statistical}.}
\label{fig:OE-statistical}
\end{figure}

In addition, experiments in Ref. \cite{mangini2022statistical} also confirmed that thermal equilibrium is approached at the fiber output when varying the input power (or the number of photons of the beam $N$), while keeping constant $u$. In particular, in Fig. \ref{fig:OE-statistical} we illustrate the process of thermalization that accompanies BSC of 7.6 ps pulses, injected into a 2 m long GRIN fiber. For clarity, and in agreement with Fig. \ref{fig:GRIN-RJ} and \ref{fig:sim-thermalization}, in Fig. \ref{fig:OE-statistical} we show the mode power fraction within each group $N_q/N$, i.e., the power in each group, divided by the group degeneracy. 

Specifically, Fig. \ref{fig:OE-statistical}a shows the evolution of the fundamental mode power fraction $N_0/N$ vs. input peak power ($\mathcal{P}$). As it can be seen, at the lowest power $\mathcal{P}$ = 0.1 kW, the fundamental mode is poorly populated. Correspondingly, the beam at the fiber output remains speckled (cfr. \ref{fig:OE-statistical}b). As the input power grows larger the power fraction of the fundamental mode increases, until eventually it acquires more than 60 \% of the total power. This nonlinear mode power redistribution is accompanied by the appearance of a bell-shape at the fiber output (cfr. Fig. \ref{fig:OE-statistical}c), i.e., the BSC effect. 
At variance with this nonlinear change of mode occupancy, $u=-U/\mathcal{P}=H/N$ and $\Omega$ remain constant, as shown in Fig. \ref{fig:OE-statistical}d. In particular, it was found that $\Omega$ = 0, in agreement with the assumption made in Sec. \ref{sec:theory}. Finally, Fig. \ref{fig:OE-statistical}e-g compares the experimental retrieved mode power fractions (histogram bars) with the thermal RJ distribution (red line) for three values of $\mathcal{P}$, i.e., 0.11, 4.64, and 8.11 kW, respectively. As it can be seen, a very good agreement between theory and experiments was found at the highest power, which corresponds to a temperature $T$ = 0.12 mm${}^{-1}$. To the contrary, whenever the power is too low, the beam does not reach thermal equilibrium upon its propagation, thus its associated mode distribution cannot be described by the RJ law.

\subsection{Thermalization by varying input coupling conditions}
\label{sec:exp-varying-H}

Baudin and co-workers \cite{baudin2020classical} carried out a different experimental study of wave thermalization in a GRIN MMF. In their experiments, thermalization was obtained in a fiber of finite length by varying the number of excited modes, as determined by the input coupling conditions, as measured by $H$ (or $U$), while operating with fixed $N$ (or input power $\mathcal{P}$). Therefore, although leading to wave thermalization, these experiments are inherently unable to describe BSC.

Their experiments were conducted with a 12 m long graded-index MMF, and using 400 ps laser pulses at 1064 nm. In order to provide a statistical relevance to their measurements, i.e., to properly place the experimental results in a statistical mechanics framework, 1000 different realizations, i.e, input conditions, were imposed for each value of $H$. Obtaining such a large set of input conditions with exactly the same value of $H$ is not trivial: it was obtained by rotating diffuser plate, which was placed right before the injection of light into the fiber core. 

Generally speaking, the near-field intensity at the fiber output can be written as the sum of two contributions, i.e., that due to the fundamental mode, that was called \textit{condensate fraction} ($I_{\mathrm{NF}}^{cond}$), and that given by the HOMs, called \textit{incoherent contribution} ($I_{\mathrm{NF}}^{inc}$).
By expressing the fiber modes within the Hermite-Gauss base, it is straightforward to derive the following expressions for $I_{\mathrm{NF}}^{cond}$ and $I_{\mathrm{NF}}^{inc}$ vs. the radial coordinate $r$:
\begin{equation}
    I_{\mathrm{NF}}^{cond} = n_0^2 r_0^2 w_0^2 \left(\frac{r}{r_0} \right),
    \label{eq:picozzi_NF_cond}
\end{equation}
\begin{equation}
    I_{\mathrm{NF}}^{inc} = \frac{1}{r_0^2} \sum_{i=1}^M n_i w_i^2 \left(\frac{r}{r_0} \right),
    \label{eq:picozzi_NF_inc}
\end{equation}
where $r_0$ is the radius of the fundamental mode, respectively, while $w_i$ is the normalized $i$-th mode of the Hermite-Gauss base, which is invariant under Fourier transform.
Analogous expressions also hold the far-field intensity profile (see Ref. \cite{baudin2020classical}).

\begin{figure}[ht!]
\centering\includegraphics[width=14.5cm]{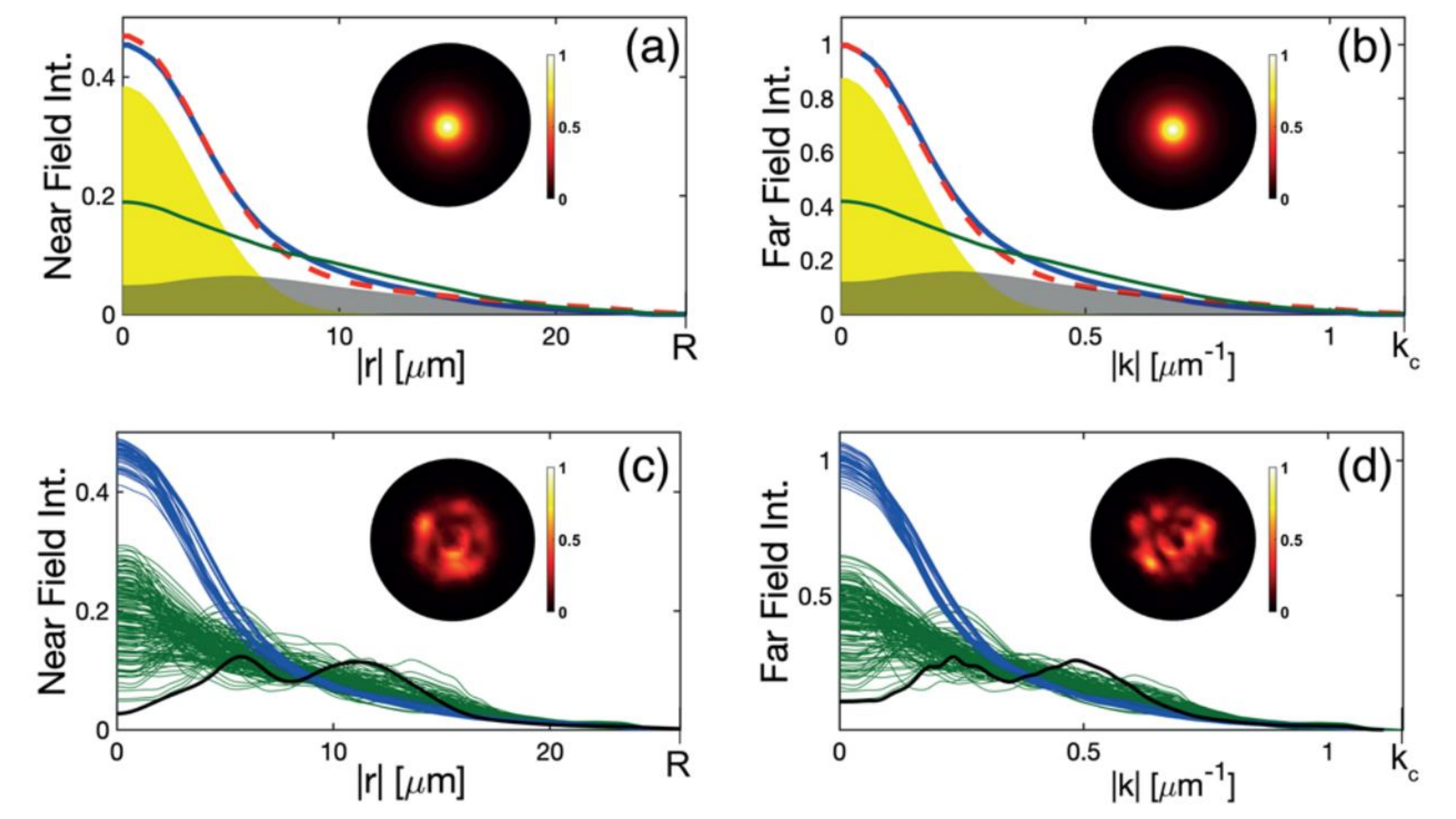}
\caption{Experimental demonstration of wave thermalization over a finite fiber length, obtained  by varying the input coupling conditions at constant input power. (a,b) Near- (a) and far-field (b) intensity profiles after 12 m of GRIN MMF. The blue line corresponds to the beam profile averaged over 1000 realizations, i.e., experiments carried out with injection conditions that provide the same value of $H$: the resulting fundamental mode power fraction is equal to 40 \%.  Yellow and gray regions in (a) correspond to the values of $I_{\mathrm{NF}}^{cond}$ and $I_{\mathrm{NF}}^{inc}$, calculated from Eqs. (\ref{eq:picozzi_NF_cond}) and (\ref{eq:picozzi_NF_inc}), respectively. The sum of $I_{\mathrm{NF}}^{cond}$ and $I_{\mathrm{NF}}^{inc}$ is shown as a red dashed line. Yellow and gray region, and red dashed line in (b) are the analogous of those in (a), but for the far-field (see Ref. \cite{baudin2020classical} for details). The inset images in (a) and (b) show output intensity distributions, averaged over all realizations. Green lines show the averaged intensity profiles recorded at only 20 cm of beam propagation. (c)–(d) Individual realizations associated with the green blue lines in (a) and (b), respectively. Insets report the intensity profile corresponding to the black lines. [Reproduced with permission from \cite{baudin2020classical}, [Baudin, Kilian, et al., Physical Review Letters 125.24 (2020): 244101.]].
}
\label{fig:exp_baudin}
\end{figure}

Experiments reported in \cite{baudin2020classical} were in excellent agreement with the theoretical predictions, showing that the averaged observed mode field profile at the fiber output is close to that predicted by the RJ law (cfr. Fig. \ref{fig:exp_baudin}). Indeed, the experimentally measured near-field intensity profile (blue line in Fig. \ref{fig:exp_baudin}a) is very close to the theoretical curve for $I_{\mathrm{NF}}$ vs. $r$ (red dashed line in Fig. \ref{fig:exp_baudin}). The latter is calculated as the sum of two contributions, i.e., from the fundamental mode, $I_\mathrm{NF}^{cond}$ (yellow area in Fig. \ref{fig:exp_baudin}a) and from the HOMs, $I_\mathrm{NF}^{inc}$ (gray area in Fig. \ref{fig:exp_baudin}a). The former was found by imposing $n_0/N$ = 0.4 in (\ref{eq:picozzi_NF_cond}). Whereas, the latter was obtained by replacing Eq. (\ref{eq:RJ-ab}) in (\ref{eq:picozzi_NF_inc}). Similar results were found by analyzing the far-field profile (see Fig. \ref{fig:exp_baudin}b).

The blue curves in Fig. \ref{fig:exp_baudin}a,b were calculated from an average over 1000 different input diffuser positions. Some of the latter are reported as blue lines in Fig. \ref{fig:exp_baudin}c,d. Moreover, in Fig. \ref{fig:exp_baudin}a,b the green curves indicate the intensity profile averaged over 1000 realizations, but obtained after only 10 cm of beam propagation. The short propagation distance measurements in Fig. \ref{fig:exp_baudin}a,b were performed in cut-back experiments, whose single realizations are shown as green curves in Fig. \ref{fig:exp_baudin}c,d, respectively. These results demonstrate that the beam thermalizes along the propagation distance, as opposed to BSC, where wave thermalization is obtained by changing the input power.

Experimental thermodynamic parameters associated with the beam at thermal equilibrium were found by fitting the experimental data, e.g., the profile of either $I_{\mathrm{NF}}$, with the sum of $I_{\mathrm{NF}}^{cond}$ and $I_{\mathrm{NF}}^{inc}$ obtained from (\ref{eq:picozzi_NF_cond}) and (\ref{eq:picozzi_NF_inc}), respectively, when considering the mode occupancy following the RJ distribution. In this way, it was possible to verify the agreement between experiments and theory as far as the dependence of $T$ and $\mu$ on $H$ is concerned. In particular, it was found that, for $H$ above a certain threshold, or equivalently \textcolor{red}{$u>u_c$}
, the fundamental mode remains as the only significantly populated mode, in agreement with the numerical predictions in Fig. \ref{fig:sim-condensation}. In conclusion, these studies can be seen as providing a formal theoretical framework for describing the experimentally well-know property that, once that only the fundamental mode of a GRIN fiber is excited at the fiber input, this mode may remain stable and uncoupled to HOMs over relatively long distances. 

\subsection{Prediction of thermodynamics parameters}

The method for determining thermodynamic parameters such as $T$ and $\mu$ depends on the experimental approach for studying wave thermalization. Specifically, for the experiments of Sec. \ref{sec:exp-varying-H}, temperature and chemical potential are retrieved by fitting the shape of the output beam profile at thermal equilibrium, after averaging over several realizations. The fitting curve is obtained by substituting the RJ distribution (\ref{eq:RJ-ok}) in (\ref{eq:picozzi_NF_inc}). 
The other way around, if the output mode power fraction is determined by means of holographic techniques, as done in the experiments of Sec. \ref{sec:exp-varying-N}, $T$ and $\mu$ can be directly computed, without recurring to any fitting procedures.
Indeed, by knowing the mode power fractions, one can determine the values of $H$ and $N$ or, equivalently, $U$ and $\mathcal{P}$. Therefore, by combining the equation of state (\ref{eq:state}), the definitions of $\mathcal{P}$ and $U$, i.e., (\ref{eq:P-ci}) and (\ref{eq:U-ci}), and the RJ law (\ref{eq:RJ-ok}), one obtains the following nonlinear equation for $T$
\begin{equation}
    \mathcal{P}=-\sum_{i=1}^M\frac{T}{\beta_i+(U-MT)/\mathcal{P}}.
\label{eq:calc-T-NL}
\end{equation}
Such an equation has one and only one physically acceptable solution, which ensures that the mode occupancy takes positive values for all modes \cite{wu2019thermodynamic}. Once that $T$ is determined, $\mu$ can be easily calculated by using the equation of state (\ref{eq:state}).

Note that, being $\mathcal{P}$ and $U$ constant during beam propagation, the values of $T$ and $\mu$ can be calculated by knowing the mode power fractions at any point along the fiber, even at the fiber input! As discussed in Sec. \ref{sec:theory}, $T$ and $\mu$ are fixed by the coupling conditions between the laser beam and the fiber. As a matter of fact, the theoretical RJ distribution at high powers in Fig. \ref{fig:OE-statistical}g was calculated by starting from the MD data at low powers, i.e., those in Fig. \ref{fig:OE-statistical}e.
This means that, although temperature and the chemical potential have a proper thermodynamic meaning at thermal equilibrium only, one may associate values of $T$ and $\mu$ to nonequilibrium states as well.

As a side note, we emphasize that a more complex algebraic problem is obtained when considering laser beams carrying OAM. In this case, an additional unknown has to be determined, i.e., the Lagrange's multiplier associated with the conservation of the OAM. This leads to a system of two nonlinear equations in two unknowns, which reads
\begin{equation}
    \begin{cases}
      \mathcal{P}=-\sum_{i=1}^M\big[\big(\frac{U}{T}+\frac{U}{T_L}-M)/\mathcal{P}+\frac{\beta_i}{T}-\frac{m_i}{T_L}\big]^{-1}\\
      U=-\sum_{i=1}^M\beta_i\big[\big(\frac{U}{T}+\frac{U}{T_L}-M)/\mathcal{P}+\frac{\beta_i}{T}-\frac{m_i}{T_L}\big]^{-1}
    \end{cases}\,,
    \label{eq:calc-T-NL2}
\end{equation}
where $T_L$ is a sort of ``angular temperature", which is associated to the Lagrange's multiplier $c$ in (\ref{eq:RJ-generalized-ab}). As it occurs for $\Omega$ = 0 even in this case there is an unique set of thermodynamic parameters that correspond to
given values of $\mathcal{P}$, $U$, and $\Omega$. 
From Eqs. (\ref{eq:calc-T-NL}) and (\ref{eq:calc-T-NL2}), it can be seen that a suitable number M of modes must be taken into consideration for the calculation of the beam temperature. In Appendix B, we discuss the error associated with calculating the value of $T$ and $\mu$ when a finite mode truncation is used, owing to practical experimental limitations.

\subsection{The role of time in experiments}

The thermodynamic approach to BSC only involves spatial properties of multimode beams. Nevertheless, in its practical demonstrations, so far BSC has been observed by using ultrashort and intense laser pulses. This incongruity, however, is only apparent. As a matter of fact, although the time dimension is not included in the thermodynamic theory, its role in experiments is crucial. Time, in fact, allows for ensuring the averaging process which is at the basis of the statistical mechanics. 
As a matter of fact, time-resolved measurements of BSC have shown that the output beam profile significantly differs between the peak and the tails of a laser pulse. Consider, in particular, the results of Ref. \cite{leventoux20213d}, where it was shown that the bell-like shape at the fiber output is the result of a time-average carried out by the camera. 

In this regard, it is worth noticing that experiments 
reported in Sec. \ref{sec:exp-varying-H} were carried out by averaging the beam output profile over many realizations, i.e., over several input conditions, that are all associated with the same thermodynamic parameters. In this way, it was possible to retrieve the output thermal equilibrium RJ mode distribution. On the other hand, in experiments carried out with ultra-short pulses, such as those described in Sec. \ref{sec:exp-varying-N}, averaging over different realizations is replaced by temporal averaging by the camera.
Accordingly, for given injection conditions, thermodynamic parameters are expected to be independent of the duration of the laser pulses. This has been experimentally verified in \cite{mangini2022statistical} by using pulses with input temporal durations ranging from 174 fs up to 7.6 ps.
In this regard, we emphasize that within state-of-the-art MD tools, carrying out an experiment for decomposing the mode content of a single laser pulse is virtually impossible. At a first sight, in fact, one may think that lowering the laser repetition rate could be a practical approach. Nevertheless, currently available holographic MD tools only permit to measure the phase and amplitude of single mode of the fiber at once; in other words, the measurement of different modes is a slow process that must be carried out in a serial fashion. 

\subsection{The role of linear disorder and Kerr nonlinearity}
\label{sec:disorder}


When considering the physical effects that are responsible for mixing power within each mode group in a MMF, inter-modal FWM (IM-FWM) is the most effective process at high powers \cite{poletti2008description}: this corresponds to the cases discussed in Sec. \ref{sec:numerical_model} and in Sec. \ref{sec:experiments}. 

Linear random mode coupling RMC \cite{ho2013linear}, which is induced by fiber bends, microbends and imperfections, is also responsible of power mixing among modes that belong to the same degenerate group, as well as, for sufficiently long propagation distances, between nondegenerate modes. Clearly, RMC is the main source of mode scrambling at low powers, where FWM effects are negligible. Here with RMC we shall just refer, for simplicity, to random coupling between adjacent nondegenerate modes. In a purely linear propagation regime, RMC alone is able to produce, similarly to FWM, a steady-state modal distribution.

For pulses propagating in a relatively long fiber span of length $L$, in the presence of Kerr nonlinearity, chromatic dispersion, linear losses and RMC, it is convenient to define some characteristic length scales. For this purpose, let us introduce an effective length $L_{\mathrm{eff}}$, over which nonlinearity is acting for a short dispersive pulse, in the presence of both fiber loss and chromatic dispersion: this distance reads as

\begin{equation}
   L_{\mathrm{eff}}=\int_{0}^{L} \frac{\exp{(-\alpha z)}}{\big[1+\big(\frac{z}{L_D}\big)^2]^{1/2}} \,dz \simeq L_D 
   \mathrm{asinh}\left(\frac{L}{L_D} \right).
\label{eq:Leff}
\end{equation}
Here $L_D$ is the distance where chromatic dispersion induces a pulse broadening of a factor $\sqrt{2}$ \cite{agrawal2000nonlinear}, and $\alpha$ (m$^{-1}$) the fiber loss coefficient. Note that in the previous definition we suppose that the nonlinearity remains much weaker than dispersion, e.g., we exclude the possible compensation of anomalous dispersion by self-phase modulation which leads to optical solitons. On the other hand, the nonlinearity length is defined as $L_{\mathrm{NL}}=1/\gamma \mathcal{P}$, with $\gamma$ (W km)$^{-1}$ the Kerr nonlinear coefficient and $\mathcal{P}$ the input pulse peak power: it is the distance over which Kerr nonlinearity produces a 1 rad phase shift. The RMC correlation length $L_{\mathrm{corr}}$ is the characteristic length, over which linear disorder leads to significant mode scrambling among adjacent nondegenerate modes. The following noteworthy scenarios are possible:

1) Strong nonlinearity, weak disorder regime: $L<<(L_{\mathrm{RMC}}$, $L_{\mathrm{eff}})$ and $L\simeq L_{\mathrm{NL}}$: RMC (among nondegenerate modes) is negligible, and FWM acts over the entire fiber span in order to redistribute energy among the fiber modes. This scenario, which was numerically simulated in Ref.\cite{sidelnikov2019random}, holds for the experiments described in Section 1 and 3.1:
the RJ distribution predicts well the equilibrium mode power distribution at the fiber output.

2) Strong nonlinearity and disorder regime: $L>>L_{\mathrm{RMC}}$, $L<<L_{\mathrm{eff}}$ and $L\simeq L_{\mathrm{NL}}$: nonlinearity acts over the entire fiber span, in concert with nondegenerate mode mixing introduced by RMC. In this scenario, an accelerated thermalization may even occur towards the RJ mode power distribution \cite{fusaro2019dramatic,sidelnikov2019random}.

3) Weak nonlinearity, strong disorder regime: $L>>(L_{\mathrm{RMC}}, L_{\mathrm{eff}})$: nonlinearity, if present, only acts over the initial section of the fiber; in the remaining part, linear RMC prevails, and it is responsible for modal power scrambling. In the limit of a purely linear regime, well-known coupled power models \cite{gloge1972optical,savovic2019power} predict an exponentially decreasing steady-state distribution of the modal power fractions $|c_i'|^2$ after a distance $L>>L_{\mathrm{RMC}}$, as well as a parabolic distribution for the mode-group powers.

\begin{figure}[ht!]
\centering\includegraphics[width=14cm]{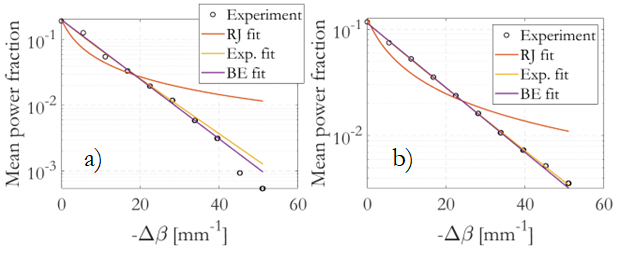}
\caption{(a) Mean power fractions for test with 1.4 ps, 5 km, 100 pJ. (b) Test with 70 fs, 830 m, 100 pJ. Data are from Ref. \cite{zitelli2023spatiotemporal}.}
\label{fig:Disorder}
\end{figure}

Fig. \ref{fig:Disorder} provides an experimental example of mode power distribution obtained within the third scenario \cite{zitelli2023spatiotemporal}. In a first experiment (Fig. \ref{fig:Disorder}a), optical pulses at 1550 nm wavelength, 1.4 ps pulsewidth and 100 pJ energy were launched into long spans (5 km)
of GRIN MMF (50/125 $\mu$m OM4 type); at the output, pulses are broadened to 285 ps by the chromatic dispersion; at the same time, modal groups are separated up to 9 ns by modal dispersion. Hence, the power $P_q$ of the individual mode groups can be easily measured by a fast photodiode, after they have experienced linear and nonlinear interactions when temporally overlapped. The mean power fraction of the modes into groups, over two polarizations, is calculated by $|c_q'|^2=N_q/N = 2P_q/g_q\mathcal{P}$, being $g_q=2, 4, .., 2Q$ the group degeneracy (including polarization). In the experiment, the effective length where nonlinearity is significant is $L_{eff}=140$ m; for the remaining 4860 m, RMC is the only effect which is responsible for modal power scrambling.
Figure \ref{fig:Disorder}a provides the average modal power fraction $|c_q'|^2$ 
vs. the differential modal eigenmodes $-\Delta\beta=\beta_q$. Fits of the experimental data are performed using the RJ law, an exponential law of the type $|c_q'|^2=a\exp{(-b\beta_q)}$, and by the Bose-Einstein distribution (BE)

\begin{equation}
   |c_q'|^2=\frac{1}{\exp{\big[-\frac{\mu+\beta_q}{T}\big]}-1}.
\label{eq:BE}
\end{equation}

Eq. (\ref{eq:BE}) represents an alternative derivation of the thermalization problem, as outlined in Section \ref{sec:subsecMatRJ}, 
which holds at low power. In Fig. \ref{fig:Disorder}a the mean modal power fraction in the linear regime is better fitted by the BE law, up to the 8-th measured modal group.

In a second experiment (Fig. \ref{fig:Disorder}b), 70 fs pulses at 1550 nm wavelength were injected into 830 m of GRIN fiber, with 100 pJ energy.  Chromatic dispersion broadens pulses to 1.1 ns; as a consequence, the effective length is only $L_{eff}=0.6$ m, and the propagation is substantially linear. Still, the figure indicates that also in this case the BE distribution properly fits the experimental data.


The example of Fig. \ref{fig:Disorder} shows that wave thermalization can be described in terms of different mode power distributions, i.e., the BE or the RJ law, according to the propagation scenario. Linear and pseudo-linear pulse propagation over long distances, dominated by RMC, lead to the BE output mode power distribution. Whereas a mode power distribution best fitted by the RJ law is observed over relatively short GRIN fiber spans, or when nonlinearity prevails over RMC.

\begin{figure}[ht!]
\centering\includegraphics[width=12cm]{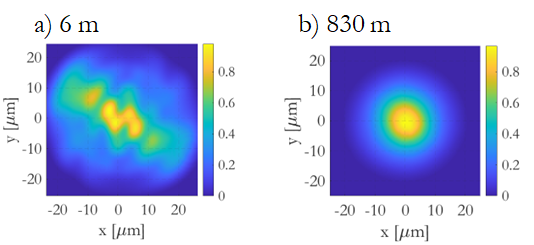}
\caption{Experimental near-field in the experiment with 70 fs, 830 m, 100 pJ, after (a) 6 m and (b) 830 m. Data are from Ref. \cite{zitelli2023spatiotemporal}.}
\label{fig:Disorder2}
\end{figure}

The BE modal distribution in Fig. \ref{fig:Disorder}b produces, in terms of near-field, a significant increase of the beam brightness, as it is illustrated by Fig. \ref{fig:Disorder2}b measured after 830 m, with respect to the highly speckled near-field of Fig. \ref{fig:Disorder2}a which measured after short distance (6 m).

\section{Conclusions, perspectives and open issues}

The wave thermalization picture provides a theoretical framework which is useful for predicting some of the key properties of BSC. Nevethless, it should be understood from the start that this appraoch has also its limitations. The most notable one being, that thremodynamics does not take into account the presence of nonlinearity, which is the very mechanism at the basis of the BSC effect!

In any case, the approach permits to design novel applications of nonlinear MMF, such as the all-optical control of the beam quality at the output of a MMF. The latter, in fact, is strictly related to the statistical temperature $T$, which can be varied via optical thermodynamic transformations, i.e., via heat exchanges among different photon gases, as it occurs for a classical gas of particles. 
Heat exchanges between photon gases in MMF have been studied only recently: for the first time, experiments of optical calorimetry in MMF have been succesfully carried out \cite{ferraro2022calorimetry}. Specifically, it was demonstrated that two photon gases which are allowed to interact by exchanging both energy and particles, eventually reach an equilibrium, i.e., the two gases have the same value of temperature and chemical potential. Moreover, thermodynamics permits to accurately predict the final value of temperature. This means that macroscopic thermodynamic variables such as the temperature and the chemical potential do not merely represent fitting parameters for an equilibrium mode power distribution, but reflect the presence of real physical forces acting at the microscopic level. Such a result, which might seem trivial when considering classical gas of particles, is instead pivotal for validating the thermodynamic theory of multimode system. As a matter of fact, it is only thanks to heat exchanges that one may ensure that the entropy is well-defined, i.e., that photon gases in MMF obeys the second law of thermodynamics. 
In this context, it is also worth mentioning that a similar concept has been recently introduced in Ref. \cite{ferraro2023multimode}, where BSC was exploited for the demonstration of an all-optical switch.

Before concluding, it is worth pointing out some of the aspects of BSC, which are not captured by the thermodynamic approach. 
For instance, BSC has been experimentally demonstrated in few-mode fibers \cite{mohammadzahery2021nonlinear}, whereas the theory only applies to highly multimode systems. Moreover, a generalized BSC effect, where a robust beam profile at the output of a MMF that is highly correlated with that of HOMs, was experimentally reported \cite{deliancourt2019kerr,deliancourt2019wavefront}. Such an effect is likely to represent an out-of-equilibrium state, in the sense that, were the fiber length much longer than in the experiment, the beam would probably decay into a mode distribution which is dominated by the fundamental mode \cite{fabert2020coherent}, as predicted by the thermodynamic theory. Nevertheless, the latter does not allow for describing BSC into other mode but the fundamental, since the thermodynamic theory only applies to equilibrium states.

Furthermore, it has to be mentioned that the analogy between nonlinear electromagnetic waves and gas of particles, which is at the basis of the thermodynamic theory, properly holds only under strong assumptions. For instance, particles are described as scalar objects, while light intrinsically requires a vectorial description. As a matter of fact, the state of polarization of light has been shown to evolve non-trivially in the process of BSC. For instance, depending on the experimental conditions, light may experience either an increase or a decrease of its degree of polarization as the input power grows larger, as reported in Ref. \cite{krupa2019nonlinear} and Ref. \cite{ferraro2022calorimetry}, respectively. Whereas, the thermodynamic theory of BSC does not explicitly consider the state of polarization of light. Only in the case of classical wave condensation, the degree of polarization is expected to increase, since only one mode should be macroscopically populated at the fiber output \cite{garnier2019wave}.

Finally, the thermodynamic theory is intrinsically an incoherent theory. That is, is not capable to describe the occurrence of any fixed phase relationship between the modes, and the improvement of the beam quality upon beam propagation. 
However, interference experiments between two spatial regions of a single self-cleaned beam, or even between two self-cleaned beams, have clearly demonstrated that the coherence of a laser beam is for the most part preserved in the process of BSC \cite{fabert2020coherent}. How to conjugate the experimental results of Ref. \cite{fabert2020coherent} and the thermodynamic theory of BSC remains, to date, an open question: we plan to address this important issue in forthcoming publications.

In conclusion, our overview has presented recent advances in the study of wave thermalization in highly multimode optical fibers. Although some important aspects, such as: the issue of beam polarization and coherence, the role of linear disorder, the study of highly nonlinear regimes, e.g., when optical solitons are formed, will require significant further work and possibly extensions of the theory beyond the realm of the present statistical theory, the thermodynamic approach has been successful in predicting the mode power distributions that correspond to self-cleaned optical beams, as well as to complex beams generated from nonlinear interactions among different individual beams. 


\section*{Acknowledgement(s)}
We are grateful to K. Krupa, A. Tonello, D. Modotto, A. Niang, D. Kharenko, M. Gervaziev, E. Podivilov, S. Babin, D. Christodoulides, F. Wise, L. Wright, F. Wu, G. Steinmeyer, G. Genty, T. Hansson, G. Millot, and A. Picozzi for many valuable discussions.

\section*{Disclosure statement}
The authors declare no conflicts of interest.

\section*{Data availability} 
Data underlying the results presented in this paper are not publicly available at this time but may be obtained from the authors upon reasonable request.

\section*{Funding}
This work was supported by the European Research Council (ERC) under the EU HORIZON2020 Research and Innovation Program (740355, 101081871), 
Ministero dell'Università e della Ricerca (R18SPB8227), 
and the European Union under the Italian National Recovery and Resilience Plan (NRRP) of NextGenerationEU, partnership on “Telecommunications of the Future” (PE00000001 - program “RESTART”).




\bibliographystyle{tfnlm}
\bibliography{interactnlmsample}

\noindent\textbf{Appendix A. Entropy calculation from experimental data}
\medskip
Among the thermodynamic parameters, the entropy $S$ is the most delicate to estimate from experimental data. At first sight, when using the definition of entropy in Eq.(\ref{eq:S-single}), i.e., $S = \sum_i \ln |c_i|^2$, one would think that the entropy diverges whenever any given mode has zero occupation. Nevertheless, this must not scare, because such a divergence only occurs because, in this case, the approximation that leads from the definition of entropy (\ref{eq:S-def-rho}) to (\ref{eq:S-ok}) is invalid. Indeed, in one recurs to the original definition of the entropy,
modes with $n_i$ = 0, i.e., $|c_i|^2 = 0$, do not contribute to 
$S$. Therefore, one can reformulate the entropy as follows:
\begin{equation}
    S = \sum_{i} S_i,
\end{equation}
where 
\begin{equation}
    S_i = \left\{
\begin{array}{lr}
   \ln |c_i|^2 & \text{if }  |c_i|^2 \neq 0\\
    0 & \text{if } |c_i|^2 = 0 
\end{array}
    \right. .
    \label{eq-Si}
\end{equation}

Experimentally, one often finds that some modes have a power fraction which is lower than the accuracy of the MD method. For instance, at thermal equilibrium, HOMs are associated with relatively power fractions when the mode group number grows larger, as predicted by the RJ distribution (\ref{eq:RJ-ok}). Therefore, in order to meaningfully evaluate the beam entropy, it is necessary to take into account the limited accuracy of the MD method on $n_i$, say $\Delta n$. In this way, one can provide an experimentally meaningful definition of the entropy as follows:
\begin{equation}
    S_i = \left\{
\begin{array}{lr}
   \ln |c_i|^2 & \text{if }  |c_i|^2 \gtrsim \Delta n\\
    0 & \text{if } |c_i|^2 \lesssim \Delta n 
\end{array}
    \right. .
\end{equation}

Finally, it must be noted that, according to its definition (\ref{eq:S-single}), $S$ grows larger with input power $\mathcal{P}$. Therefore, with the goal of studying the entropy growth due to a nonlinear mode power redistribution during the thermalization process that accompanies BSC, it is convenient to define the entropy per particle, say, $\Tilde{S}$. The latter can be determined by writing the entropy $S$ as 
\begin{equation}
    S = \Tilde{S} + M \ln \mathcal{P},
\end{equation}
where
\begin{equation}
    \Tilde{S} = \sum_{i=1}^M\ln \left(\frac{|c_i|^2}{\sum_j |c_j|^2} \right).
\end{equation}

\medskip
\bigskip
\noindent\textbf{Appendix B. Mode truncation error}
\medskip

The number of guided modes in a MMF depends on several factors, such as the laser wavelength, the geometrical properties of the fiber, and the refractive index difference between core and cladding. Generally speaking, standard GRIN fibers propagate hundreds of modes at wavelengths around 1 $\mu$m. Whereas, owing to intrinsic limitations, state-of-the-art MD techniques do not permit to resolve all of the fiber modes with sufficient precision. Therefore, a mode truncation errors inevitably affect the results of MD experiments.
In this regard, it is interesting to note that mode truncations do not affect the estimation of the chemical potential. Indeed, the latter can be computed by the bare knowledge of the ratio between the power fraction of any pair of modes. At thermal equilibrium, in fact, the mode occupancy obeys the RJ law, so that the ratio between the power fraction per mode of the $q$-th and $r$-th mode groups reads
\begin{equation}
      \frac{|c_q|^2}{|c_{r}|^2}=\frac{\mu+\beta_{r}}{\mu+\beta_q},
\end{equation}
which provides a unique value for $\mu$, provided that $r\neq q$. 

On the other hand, when estimating the temperature $T$ from MD data, mode truncation introduces a significant sources of error. In fact, at variance with $\mu$, the value of $T$ strongly depends on the number of groups of guided modes $Q-1$
. By determining $T$ from the equation of state, i.e.,
\begin{equation}
T = \frac{-\sum_q^{Q-1} g_q \beta_q |c_q|^2 -\mu\sum_q^{Q-1} g_q|c_q|^2}{\sum_q^{Q-1}g_q},
\end{equation}
it is straightforward to see that $T$ depends on $Q$
.


\end{document}